\documentstyle[aps,psfig]{revtex}
\begin{document}
\def\half{{1\over 2}}
\newcommand{\bra}[1]{\langle#1 |}
\newcommand{\ket}[1]{| #1\rangle}
\draft
\title{Fluctuation properties of strength functions associated with
giant resonances} 
\author{Hirokazu Aiba$^1$~,~ Masayuki Matsuo$^2$~,~
Shigeru Nishizaki$^3$~,~and ~Toru Suzuki$^4$}
\address{$^1$Kyoto Koka Women's College, 38 Kadono-cho Nishikyogoku, Ukyo-ku,
         615-0882 Kyoto, Japan\protect\\
     $^2$Graduate School of Science and Technology,
      Niigata University, 950-2181 Niigata, Japan\protect\\
     $^3$Faculty of Humanities and Social Sciences, 
     Iwate University, 3-18-34 Ueda, 020-8550 Morioka, Japan\protect\\
     $^4$Department of Physics, Tokyo Metropolitan University, 
     192-0397 Hachioji, Japan}
\vspace{1cm}
\date{\today}
\maketitle
\begin{abstract}
We performed fluctuation analysis by means of the local scaling
dimension for the strength function of the isoscalar (IS) and the isovector
(IV) giant quadrupole resonances (GQR)
in $^{40}$Ca, where the strength functions are obtained by the shell model
calculation within up to the 2p2h configurations. 
It is found that at small energy scale, 
fluctuation of the strength
function almost obeys the Gaussian orthogonal ensemble (GOE) random 
matrix theory limit.
On the other hand, we found a deviation from the GOE limit at the 
intermediate energy scale about 1.7MeV for the IS and at 0.9MeV for the IV.
The results imply that different types of fluctuations coexist at different 
energy scales.
Detailed analysis strongly suggests that 
GOE fluctuation at small energy scale is due to the complicated nature of
2p2h states and that fluctuation at 
the intermediate energy scale is associated
with the spreading width of the Tamm-Dancoff 1p1h states.
\end{abstract}

\section{Introduction}
\label{sec:intro}

Strengths of exciting  high lying states in nuclei, for instance,
a giant resonance, are spread over a certain energy interval
due to the coupling to a background which is complicated and has
huge degrees of freedom. As a result, the strength function, at a large
energy scale, exhibits a global shape profile like a Lorentzian.
The values of the center of energy or the total width depend on the properties
of a probe such as multiporality, isospin, and so on, and change smoothly as 
the mass number of a nucleus. Most of these values are well 
understood up to now \cite{bertsch,speth,Harak}.
On the other hand, at small energy scale limit comparable to the level
spacing of the background states, the strength function 
may rapidly fluctuate from state to state
on top of the global profile, if we neglect the escaping width due to the 
coupling to the continuum. 
It is believed that the fluctuation properties
at this energy scale can be well simulated by the random matrix theory of the
Gaussian orthogonal ensemble (GOE) \cite{dyson,mehta,bohigas,brody}, 
which has been verified by several
experimental data: The nearest-neighbor level spacing distribution (NND) of
neutron resonances near the threshold approximately follows the Wigner
distribution \cite{brody,Bohr-Mottelson1}.
The distribution of the reduced width (proportional to an absolute
square of a component of the wave function) of  these resonance
also shows the Porter-Thomas distribution \cite{brody,Bohr-Mottelson1}.
Furthermore, the NND of the Nuclear Data Ensemble shows the Wigner 
distribution and the $\Delta_3$ statistics of them shows 
$\Delta_3(L)\propto\ln L$ \cite{haq}.
All of these are typical signatures of the GOE.

Thus, we know well the behavior of the strength function  at both
extreme small and large energy scales. From this knowledge, 
one may construct the following model to describe
the strength function of  highly excited states in nuclei:
 A collective state couples democratically to each of
unperturbed background states, and the background Hamiltonian itself
is the GOE random matrix. This GOE background model is closely 
connected with the pandemonium
picture \cite{hansen,pandemonium}.
If we adjust the energy of the collective state 
and the coupling strength
between the collective state and background states, the resultant strength
function should be consistent with the giant resonance 
strength function at least
at both extreme energy scale limits. Therefore, in order to find the 
difference between a realistic strength function and that from this model,
we must investigate fluctuation properties of the strength
function at intermediate energy scale \cite{kilgus,kamer}.

Recently, we proposed a new method to perform fluctuation analysis
of the strength function \cite{aiba}. This method is devised to quantitatively 
characterize a
fluctuation property as a function of energy scale
by a new measure called the local scaling dimension (LSD), 
and is suitable to
investigate fluctuation at intermediate energy scales.
We applied this method to a simple doorway damping model.
The doorway damping model is different from the above GOE background
model with respect to two points: First, unperturbed background states
are divided into two different classes so that unperturbed states in one class,
which are called the doorway states, couple directly to a collective state,
while states in another class do not. The second point is that doorway
states and other unperturbed background states have a finite spreading width
due to the mixing among them. (Note that the GOE background model 
corresponds to the large spreading width limit.) 
Fluctuation properties of the strength functions of both models are same at 
small energy scale limit.
The result of the analysis clearly showed that fluctuation properties of
both strength functions deviate from each other at a certain intermediate
energy scale, and this energy scale is closely related to the spreading
width of the doorway states in the doorway damping model.

In this paper, we apply the same method to a more realistic strength function
of the giant resonances in a nucleus. Strength functions of the isoscalar (IS)
and the isovector (IV) giant quadrupole resonance (GQR) in $^{40}$Ca are 
obtained by the shell model
calculation within up to the 2p2h configurations.
The strength functions reproduce the experimental center of energy and
the total width, and also show the GOE fluctuation at the small
energy scale limit, as will be shown later. Furthermore, in this model,
the doorway structure  and the deviation from GOE in the background
may be  introduced in a natural way, while they are input by hand in the
doorway damping model. Therefore these strength functions may provide the better
test for the analysis.

Note that a similar method of analysis was proposed and applied to the
($e,e'$) or ($p,p'$) experimental data on $^{208}$Pb \cite{lacroix,lacroix2}.

The paper is organized as follows: In Sec.\ \ref{sec:lsd}, we briefly
explain the LSD. In Sec.\ \ref{sec:analysis}, we apply the method to
the strength functions of the isoscalar (IS) and the isovector (IV)
GQR. We also discuss the physical
origin of the deviation from the GOE background model in detail.
Finally, Sec.\ \ref{sec:conclusion} is devoted to conclusions.

\section{Local Scaling Dimension}
\label{sec:lsd}
We briefly explain the local scaling dimension in this section.
See Ref.\ \cite{aiba}  for detail.

The strength function is expressed as \cite{Bohr-Mottelson2}
\begin{equation}
S(E)=\sum_i S_i\delta(E-E_i+E_0),
\label{defstr}
\end{equation}
for exciting the nucleus with excitation energy $E$.
Here $E_i$ and $E_0$ are the energy of discrete
levels and the ground state energy, respectively,
and $S_i$ denotes the strength of exciting the $i$th energy level.
Let us assume strengths are normalized as $\sum_i S_i=1$.

In order to quantitatively characterize how the strength
function  fluctuate  at various energy scale,
we consider binned distribution of the 
strength function $S(E)$ by dividing whole energy interval under 
consideration into $L$ bins with length $\epsilon$. Strength
contained in $n$th bin is denoted by $p_n$,
\begin{equation}
p_n\equiv\sum_{i\in n{\rm th~ bin}}S_i.
\label{defp}
\end{equation}
To characterize the distribution of the binned strengths,
we introduce the moments of it, which are called in literature
the partition function $\chi_m(\epsilon)$ defined by
\begin{equation}
\chi_m(\epsilon)\equiv \sum_{n=1}^L p_n^m \\
                 =L\langle p_n^m\rangle.
\label{partition}
\end{equation}
Clearly, the partition functions  contains, at small $\epsilon$ limit,
the information of the strength distribution , as well as 
the energy-strength correlation as a function of
a bin width or an energy scale $\epsilon$.

It is then helpful to employ the idea of the generalized fractal 
dimension \cite{hentschel,halsey},
which is derived from the partition function and has a definite physical
meaning. The concept of the generalized fractal dimension is useful 
only when a system has a multifractal nature, namely,
has a self-similar structure against the change of the scale. Strength functions
at a nucleus generally do not have such a property.
Thus, we have to extend the concept of the generalized fractal dimension and 
finally reach the local scaling dimension (LSD) defined by
\begin{equation}
D_m(\epsilon)\equiv {1\over m-1}
\frac{\partial\log\chi_m(\epsilon)}
{\partial\log\epsilon}.
\label{scaledim}
\end{equation}
Different from the generalized fractal dimension, the local
scaling dimension characterizes ``how the partition function scales 
for different energy scales'' and accordingly is a function of a bin 
width or an energy scale $\epsilon$.
The local scaling dimension, however, has an easily understandable
physical meaning similar to those of the generalized fractal dimension:
For instance, the value of $D_m(\epsilon)$ close to unity  means that
the strength function looks like a one dimensional object distributed
smoothly over an energy interval when we look at the strength function
at an energy scale $\epsilon$. On the other hand, it looks like a zero 
dimensional object, namely a dot  if the value
of $D_m(\epsilon)$ is close to zero.
Note that the local scaling dimension reduces to the generalized fractal
dimension when the value of $D_m(\epsilon)$ is constant over a whole interval
of $\epsilon$.
\begin{figure}
\begin{center}
  \begin{minipage}{7.5cm}
       \psfig{file=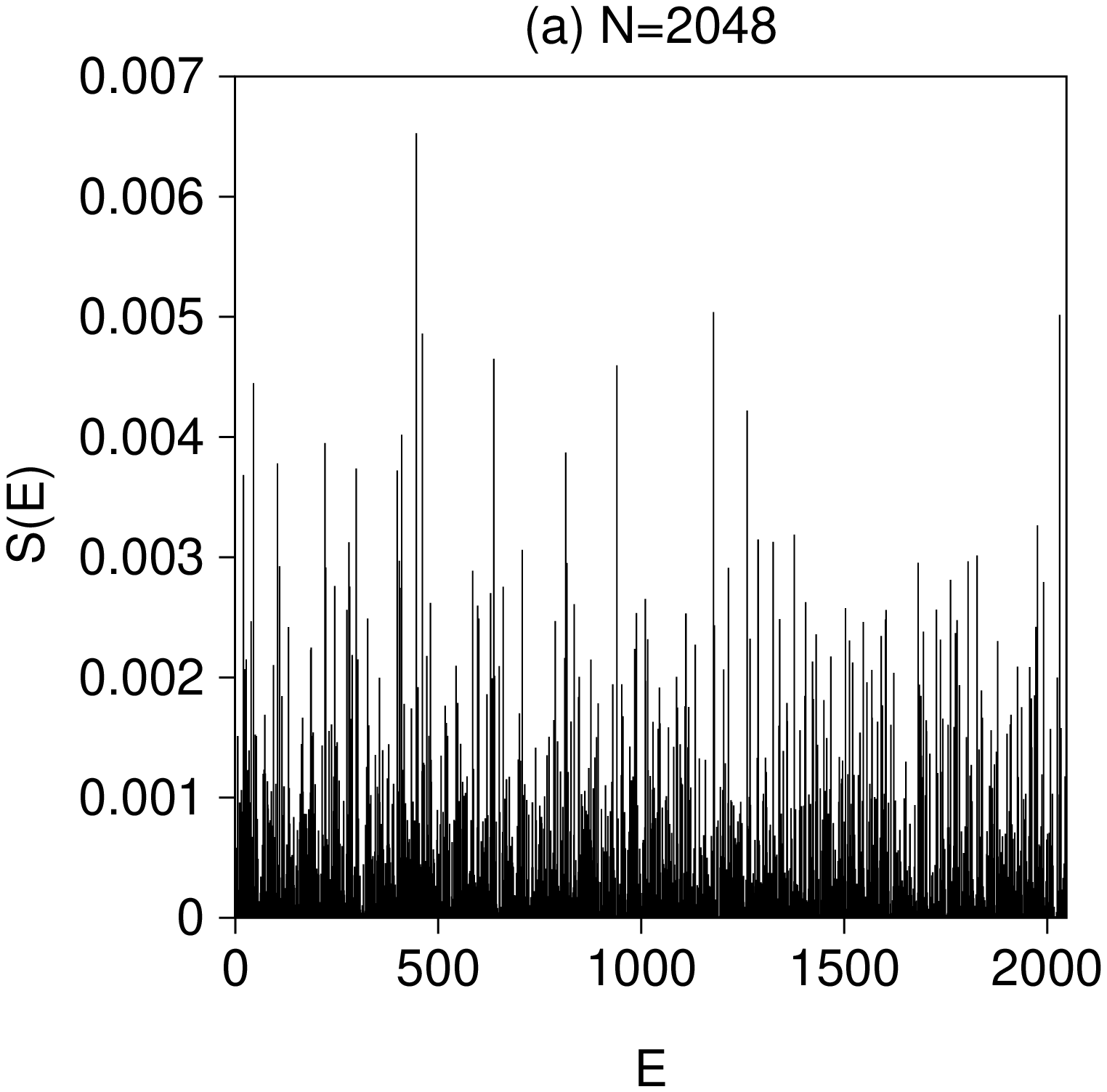,width=7.5cm}
  \end{minipage}
  \begin{minipage}{7.5cm}
       \psfig{file=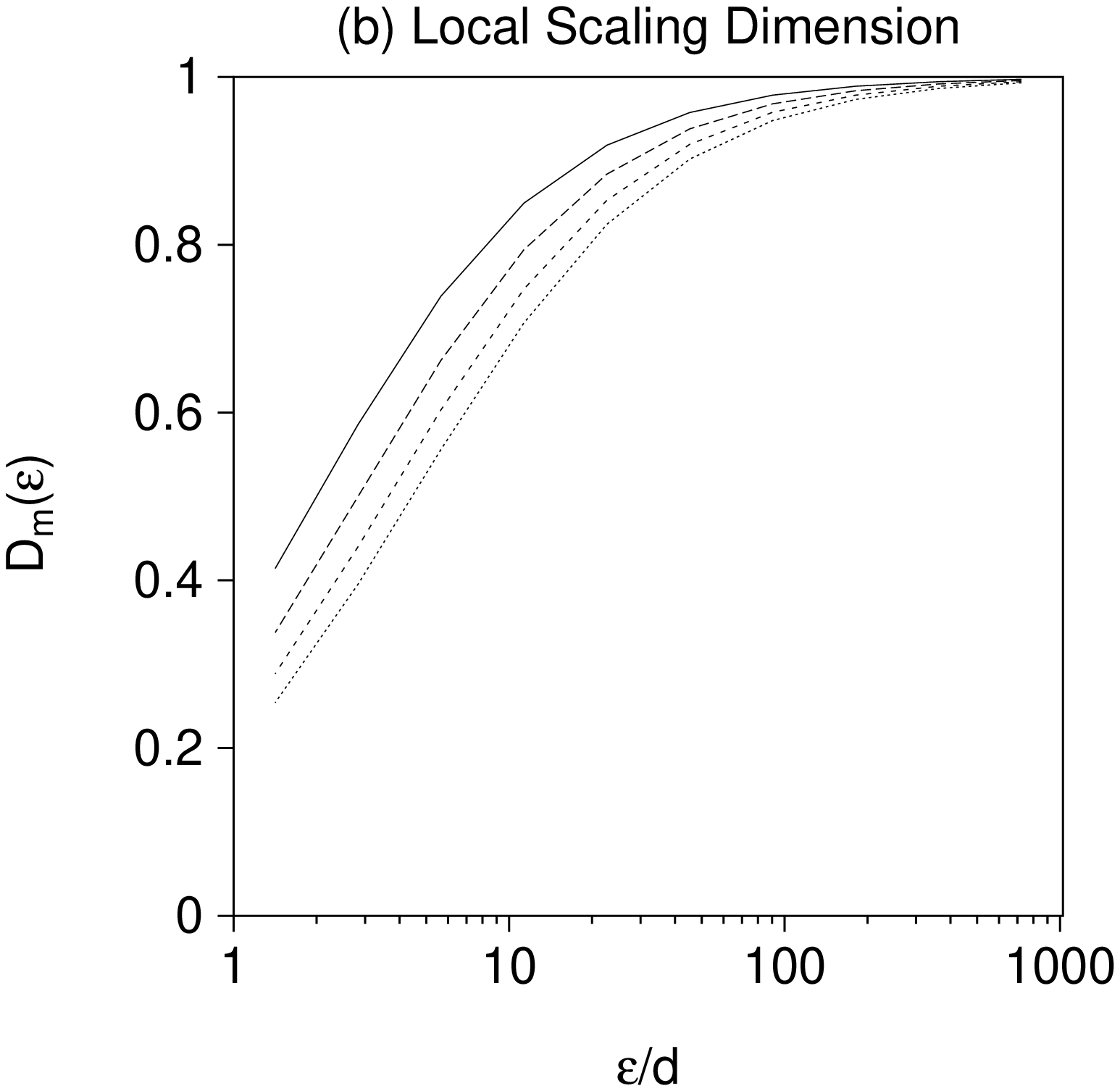,width=7.5cm}
  \end{minipage}
\end{center}
 \caption{
(a)An example of the GOE strength function with matrix dimension of 2048.
(b)The local scaling dimension $D_m(\epsilon)$ for the GOE strength
 function obtained after the ensemble average. The curves correspond to
 $D_m(\epsilon)$ for $m=2$ to 5 from upper to lower.
}
\label{fig_goe}
\end{figure} 
In actual calculation of the local scaling dimension, we 
define it by means of the finite difference under the change
of a factor 2,
\begin{equation}
D_m(\sqrt{2}\epsilon) = {1\over m-1}\frac{
\log\chi_m(2\epsilon)-
\log\chi_m(\epsilon)}
{\log 2},
\label{approscaledim}
\end{equation}
rather than the derivative in Eq.\ (\ref{scaledim}).
Using the finite difference, the calculation is very simple for 
all the moments. 

One of the important strength functions may be the one obtained from 
the GOE random matrix. The local scaling dimension of this GOE strength
function can be used as a reference to study other strength functions
which are supposed to have a similar fluctuation properties to the GOE,
as in the case of nuclear strength functions in the highly excited region.
We show in Fig.\ \ref{fig_goe} an example of the GOE strength function
and the local scaling dimension $D_m(l)$, where $l\equiv\epsilon/d$ 
(the bin width measured in the unit of the level spacing $d$) 
is used as the scaling parameter.

\section{Analysis of the giant quadrupole resonances 
in $^{40}$C\lowercase{a}}
\label{sec:analysis}
We apply the local scaling dimension to the strength functions
of the giant quadrupole resonances in $^{40}$Ca.

\subsection{Numerical Calculation of Strength Functions}
\par
  We calculated the strength functions of the isoscalar and the isovector
giant quadrupole resonances in $^{40}$Ca within the second Tamm-Dancoff
approximation including the 1p1h and 2p2h excitations. Single-particle
wave-functions and energies were obtained in terms of a Woods-Saxon potential
including the Coulomb interaction. To simulate the bare (Hartree-Fock)
single-particle energies $\varepsilon_{HF}$, the above single-particle energies
$\varepsilon_{WS}$ were scaled by the effective mass $m^*/m$
as $\varepsilon_{HF}=\varepsilon_{WS}/(m^*/m)$.
As the residual interaction, the Landau-Migdal-type interaction \cite{Schwe}
including the density-dependence was adopted.
The model space was constructed in terms of single-particle states
within the four major shells, two below and two above the Fermi surface,
and included all 1p1h states and 2p2h states whose unperturbed energies were
less than 50MeV. Resultant number of 1p1h states and 2p2h states
are 34 and 4144, respectively. We diagonalized the Hamiltonian
within this model space and obtained the strength functions for
the isoscalar and the isovector quadrupole operators.

\begin{figure}
\begin{center}
  \begin{minipage}{7.5cm}
       \psfig{file=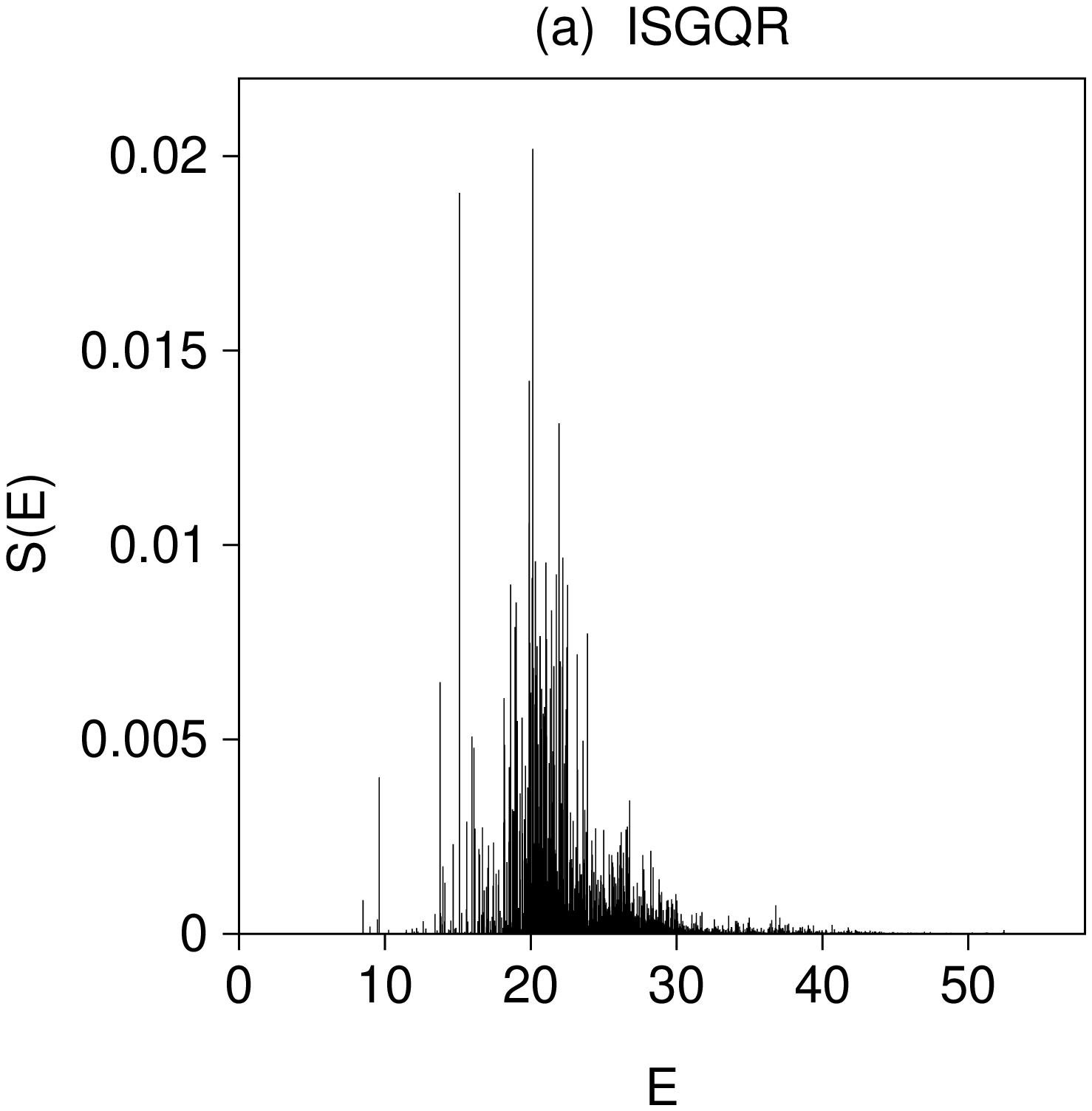,width=7.5cm}
  \end{minipage}
  \begin{minipage}{7.5cm}
       \psfig{file=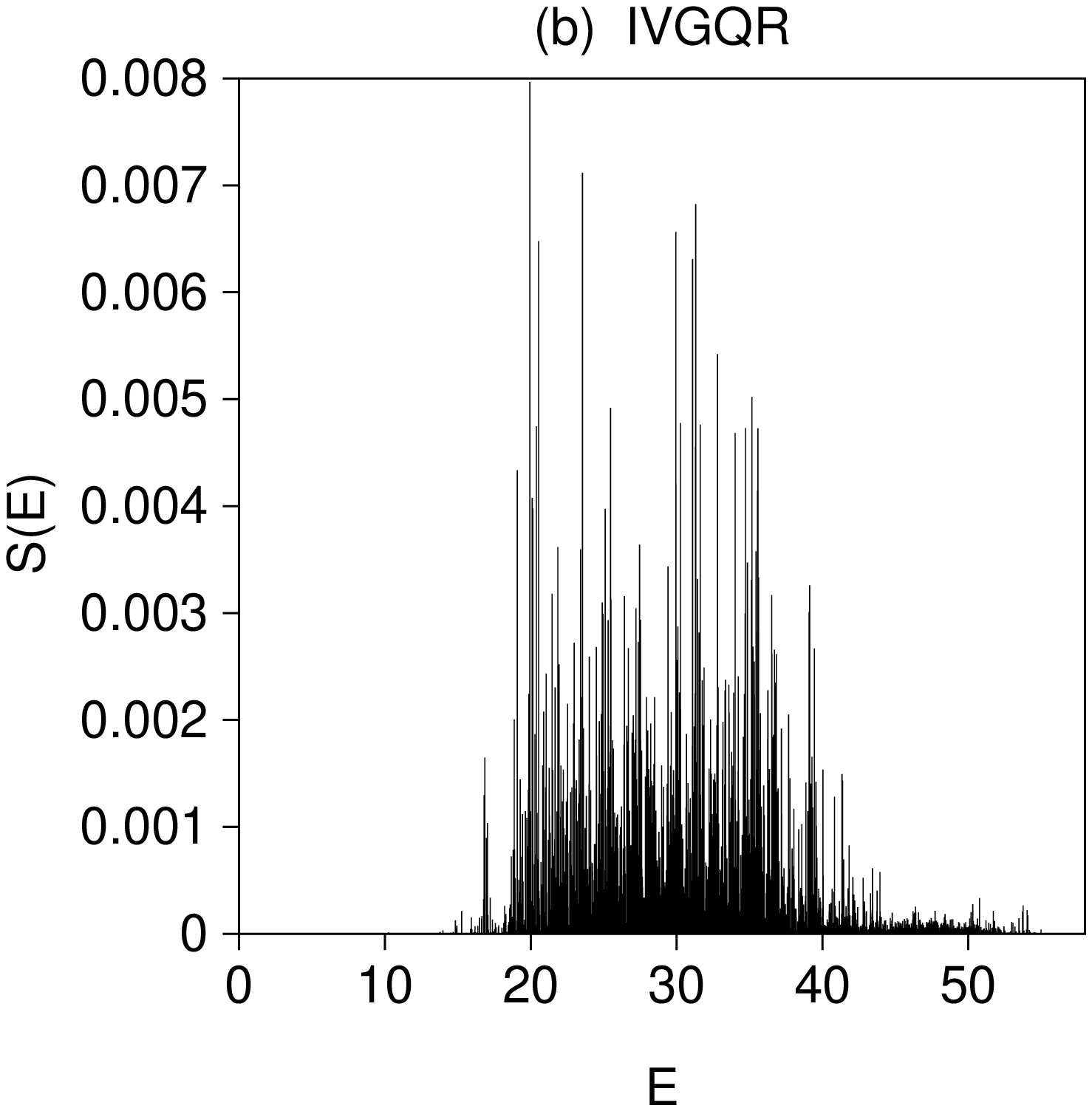,width=7.5cm}
  \end{minipage}
\end{center}
 \caption{
Strength functions for ISGQR (a) and IVGQR (b).
}
\label{fig_strfun}
\end{figure}

Figure \ \ref{fig_strfun} (a) and (b) show the strength functions for ISGQR and
IVGQR, respectively. The mean energy and the total width
(the standard deviation) are 23MeV(31MeV) and 4.0MeV(7.0MeV)
for the ISGQR(IVGQR). The effective mass $m^*/m=0.9$ was used
to reproduce the empirical mean energy for IVGQR in $^{40}$Ca \cite{Harak}.
These strength functions were already used in Refs. \cite{Droz1} and
\cite{Droz2} to study the generic properties of nuclear giant-resonance decay.

\subsection{Several Measures}

Before going to the detailed discussion for the local scaling
dimension, we briefly show other fluctuation measures.

\begin{figure}
\begin{center}
  \begin{minipage}{7.5cm}
       \psfig{file=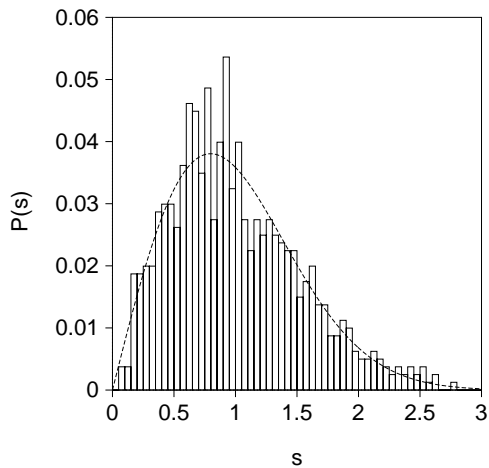,width=7.5cm}
  \end{minipage}
\end{center}
 \caption{
The nearest neighbor level spacing distribution for levels
between 20MeV and 30MeV.
Level spacings were unfolded by the Strutinsky method with
a smoothing width 5MeV. The dashed curve represents
 the Wigner distribution.
}
\label{fig_nns}
\end{figure} 

\begin{figure}
\begin{center}
  \begin{minipage}{7.5cm}
       \psfig{file=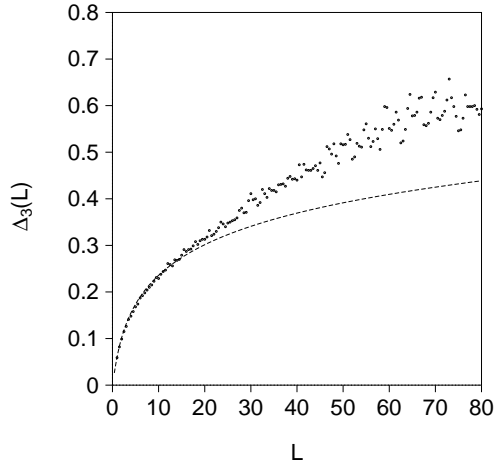,width=7.5cm}
  \end{minipage}
\end{center}
 \caption{
The $\Delta_3$ statistics. The dashed curve represents
the $\Delta_3$ for the GOE level fluctuation.
Same as Fig.\ \protect\ref{fig_nns} for others.
}
\label{fig_delta3}
\end{figure}

Figure\ \ref{fig_nns} shows the nearest-neighbor level spacing
distribution (NND), where levels between 20MeV and 30MeV are considered
(the number of levels is 804), and are unfolded by means of the Strutinsky 
method with a smoothing width 5MeV \cite{ring-schuck}.
The NND follows rather well the Wigner distribution, which indicates
at the very small energy scale the level fluctuation almost obeys the GOE.
We also plot in 
Fig.\ \ref{fig_delta3} the $\Delta_3$ statistics. We again find that at small
energy range the $\Delta_3$ follows the GOE line.
At $L^{\rm max}\simeq 15$, however, $\Delta_3$ starts to deviate from the
GOE line to upward. This value of $L^{\rm max}$ roughly 
corresponds to the energy scale $E\simeq0.18$MeV.

\begin{figure}
\begin{center}
  \begin{minipage}{7.5cm}
       \psfig{file=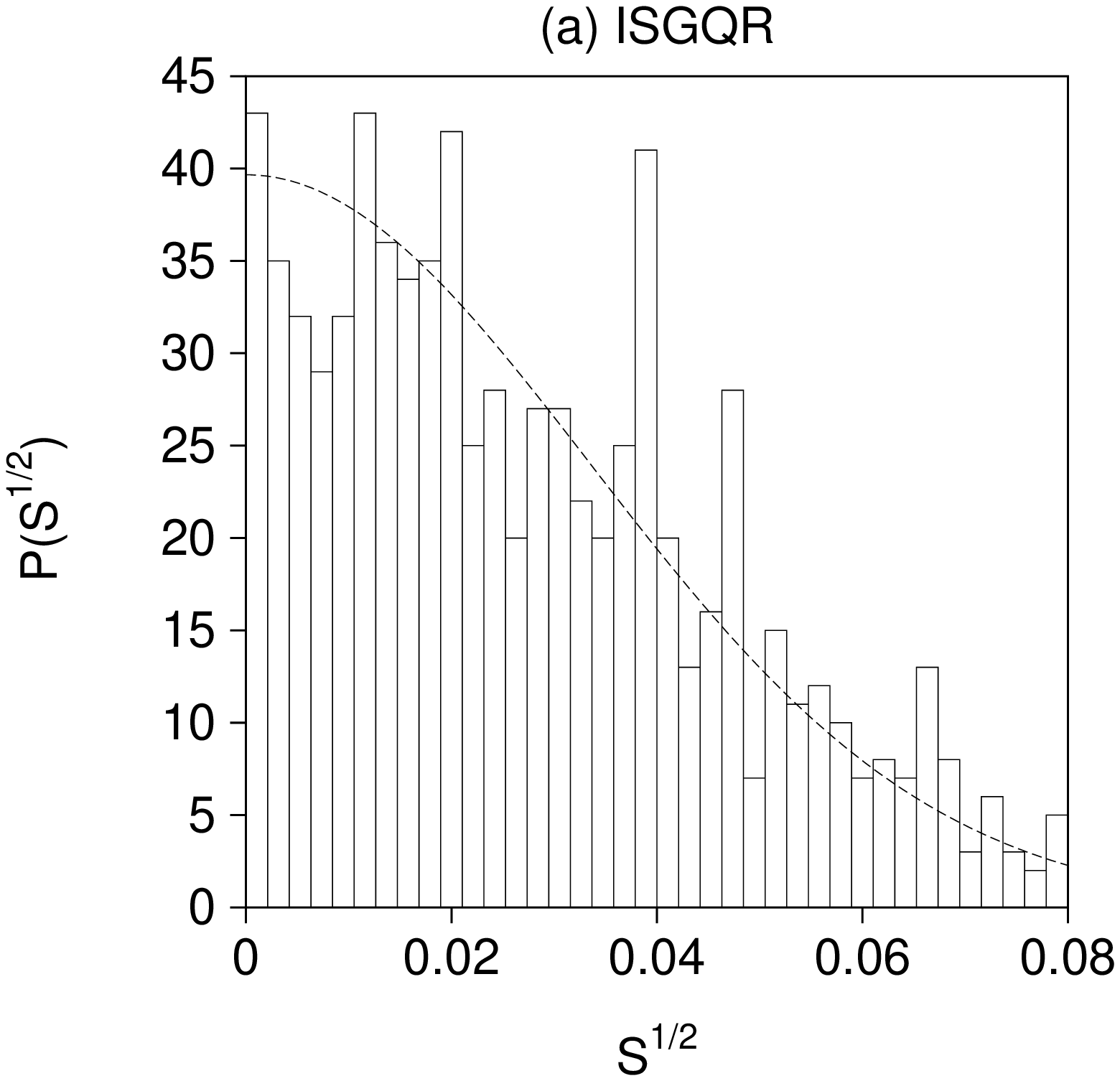,width=7.5cm}
  \end{minipage}
  \begin{minipage}{7.5cm}
       \psfig{file=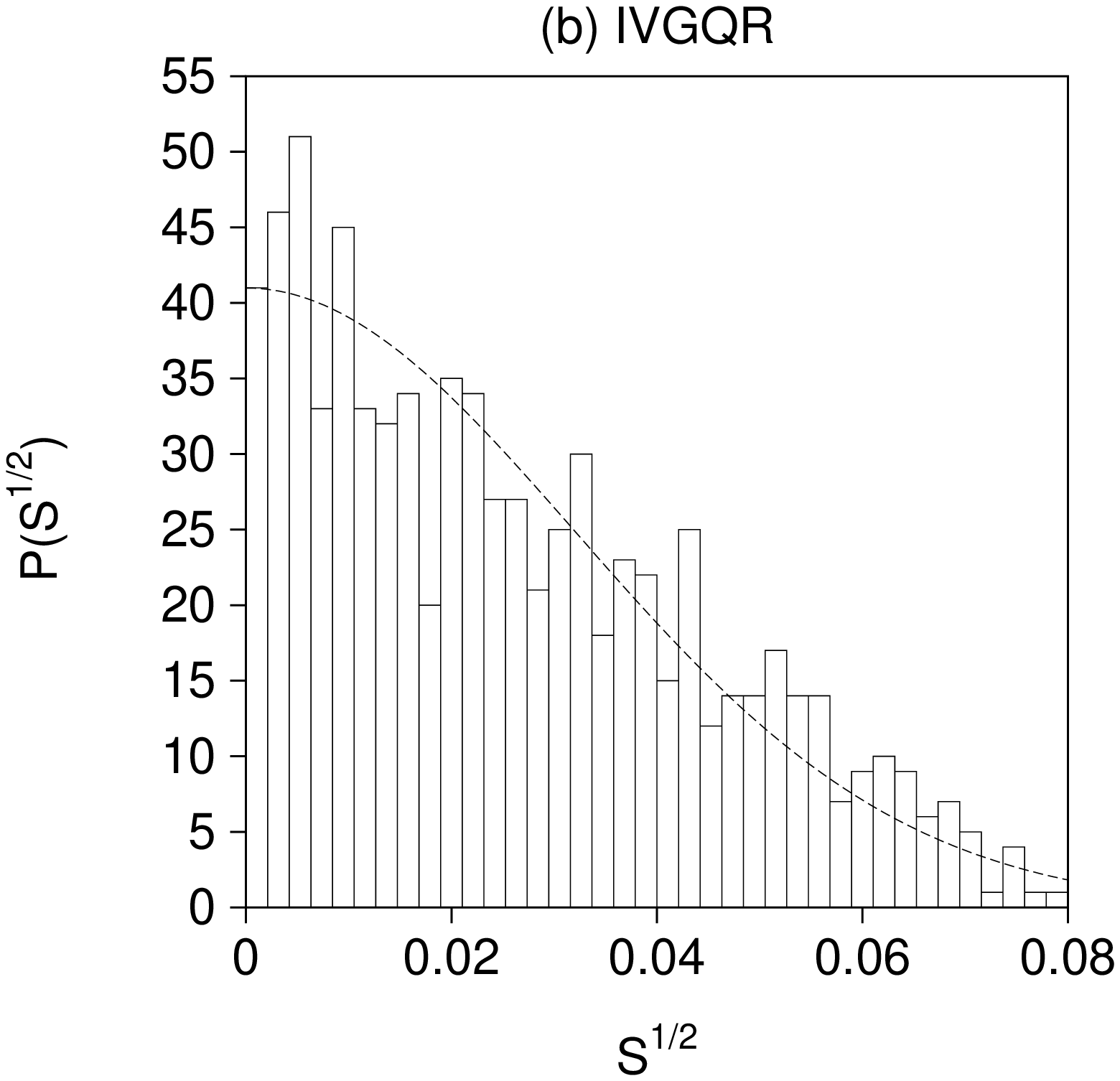,width=7.5cm}
  \end{minipage}
\end{center}
 \caption{
The statistical distribution of square root of 
unfolded strengths, $\bar{S}_i^{1/2}$,
associated with the individual levels between 20MeV and 30MeV
for (a)ISGQR and (b)IVGQR. 
 The dashed curve represents the Porter-Thomas
distribution which becomes a Gaussian when plotted as a function
of $\bar{S}_i^{1/2}$.
}
\label{fig_strdis}
\end{figure}

Figure\ \ref{fig_strdis} shows the strength distribution where the distribution
of the square-root of normalized strengths $\bar{S}_i$ is plotted
as a histogram; strengths are normalized also by Strutinsky method, detailed
procedure of which will be explained in the next subsection.
We can see that the strength 
distribution follows considerably well the Porter-Thomas distribution.


\subsection{Results of the Local Scaling Dimension}

Since we are not interested in the global shape of the strength
function, 
it is convenient to introduce the normalized strength function where
the global smooth energy dependence is removed from the original strength
function as follows:
\begin{equation}
\bar{S}_i ={\cal N}{S_i\tilde{\rho}(E_i) \over
\tilde{S}(E_i)}.
\label{eq_norstr}
\end{equation}
Here, 
$\tilde{S}(E)$ and $\tilde{\rho}(E)$ are obtained by averaging 
the calculated strength function, Eq.(\ref{defstr}),
and the level density $\rho(E)=\sum_i\delta(E-E_i+E_0)$, respectively,
by the Strutinsky method. We adopted 5MeV as a smoothing width.
$\cal{N}$ is an overall normalization factor 
to guarantee $\sum_i\bar{S}_i=1$ for the considered energy interval.
Moreover, we adopt as energy levels the equidistant ones \cite{aiba},
namely, $E_i=id$, where $d$ denotes average level spacing.
By adopting these levels, we can neglect global energy dependence
of the level density and the local level fluctuation. 
(We verified that the inclusion of the local level fluctuation
affects only the behavior of the LSD at small energy scales, and
this effect can be described by the GOE.)
Thus, we obtain the normalized strength function ${\bar S}(E)$;
\begin{equation}
{\bar S}(E)=\Sigma_i{\bar S}_i\delta(E-E_i+E_0).
\label{eq_norstrfun}
\end{equation}
The local scaling dimension is derived from this normalized strength
function.
Finally, since the visible strengths of ISGQR are almost concentrated in
the energy range between 20MeV and 30MeV, we adopt this
energy interval for analysis. 

\begin{figure}
\begin{center}
  \begin{minipage}{7.5cm}
       \psfig{file=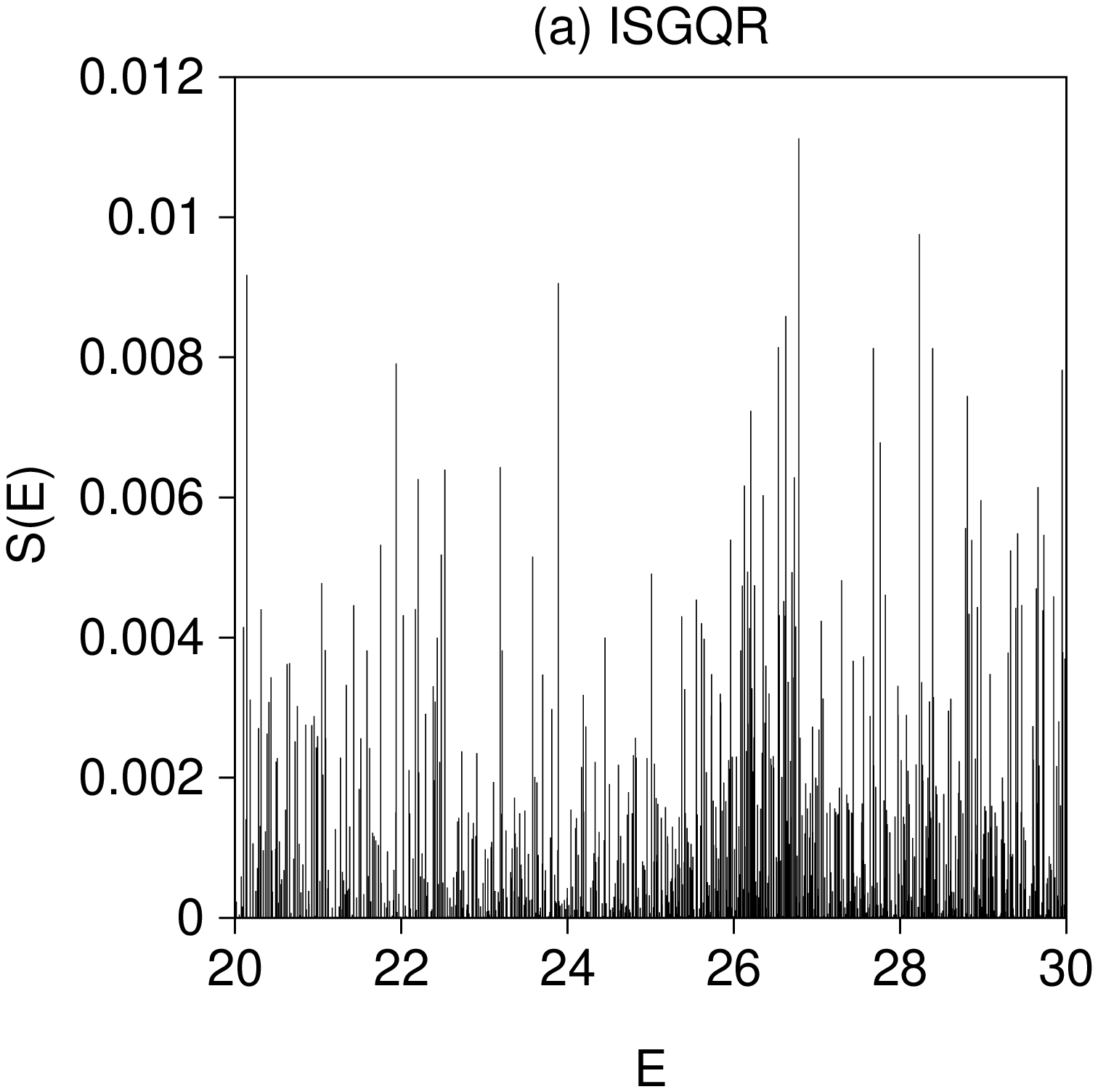,width=7.5cm}
  \end{minipage}
  \begin{minipage}{7.5cm}
       \psfig{file=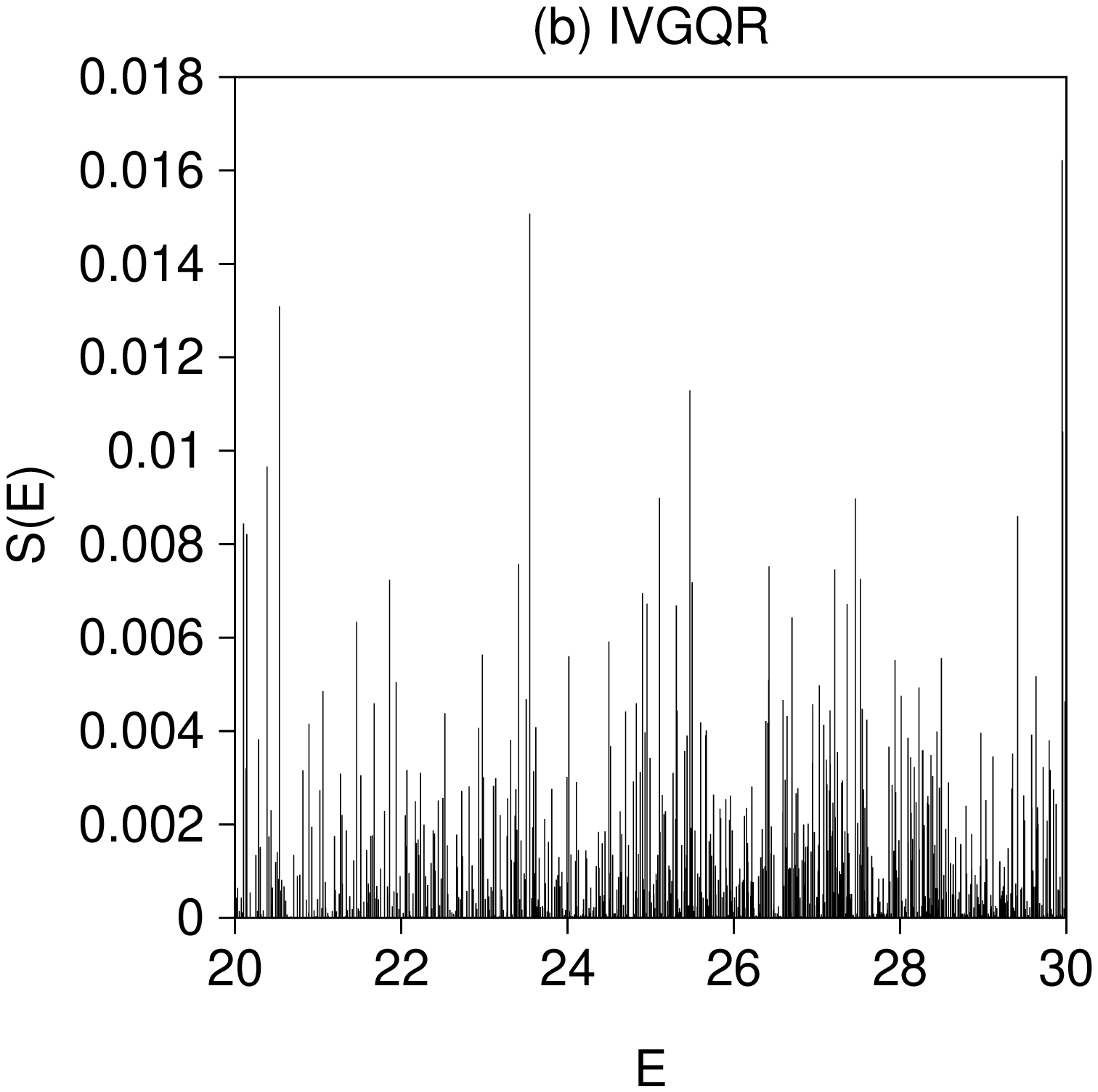,width=7.5cm}
  \end{minipage}
\end{center}
 \caption{
The normalized strength functions ${\bar S}(E)$ corresponding to the
original ones for the ISGQR (a) and for the IVGQR (b) 
plotted in Fig.\ \protect\ref{fig_strfun}. A summation of strengths within
an energy interval 20MeV-30MeV is normalized to unity.
}
\label{fig_norstrfun}
\end{figure} 

The normalized strength functions for ISGQR and IVGQR
are plotted in Figs.\ \ref{fig_norstrfun} (a) and (b), respectively.

\begin{figure}
\begin{center}
  \begin{minipage}{7.5cm}
       \psfig{file=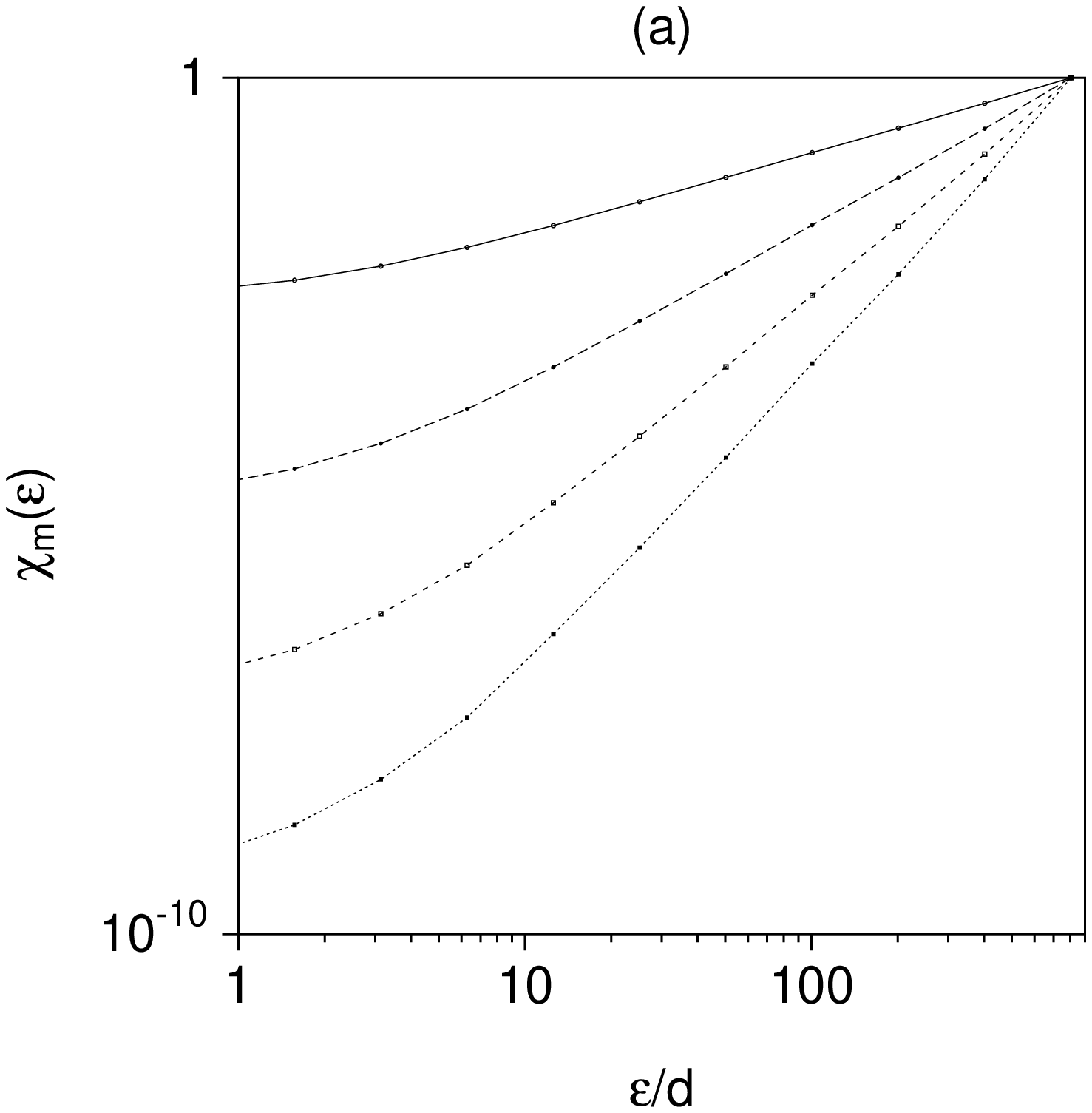,width=7.5cm}
  \end{minipage}
  \begin{minipage}{7.5cm}
       \psfig{file=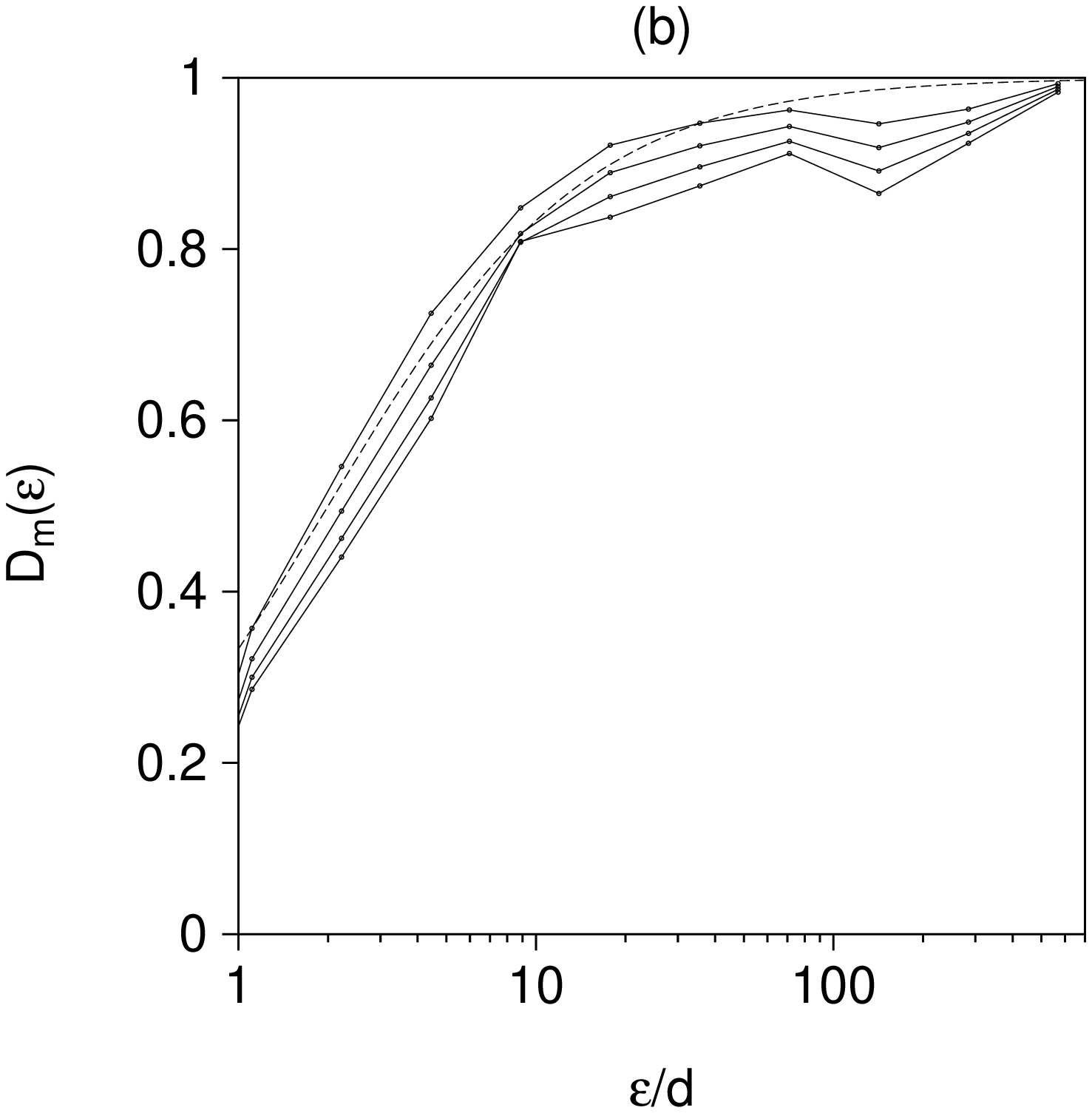,width=7.5cm}
  \end{minipage}
\end{center}
 \caption{
(a)The partition function $\chi_m(\epsilon)$ for $m=2$ to 5
for the ISGQR corresponding to
the normalized strength funciton plotted in Fig.\ \protect\ref{fig_norstrfun}
 (a). (b) Its local scaling dimension $D_m(\epsilon) $for $m=2$ to 5. 
The dashed curve represents $D_2(\epsilon)$ for the GOE.}
\label{fig_lsdis}
\end{figure}

First, let us discuss the ISGQR case in detail.
The case of the IVGQR will be discussed in Sec.\ \ref{sec:isovector}.
Figures\ \ref{fig_lsdis}(a) and (b) show the partition
functions $\chi_m(\epsilon)$ defined by Eq.\ (\ref{partition}) and the local 
scaling dimensions $D_m(\epsilon)$ defined by
Eq.\ (\ref{approscaledim}), respectively. 
The horizontal axes in both figures represent the
bin width of energy axis in unit of $d$, where
 $d$ represents the average level
spacing over the energy range 20MeV and 30MeV ($d\simeq12$keV). From 
the partition function
$\chi_m(\epsilon)$, one sees that the
fluctuation does not follow the power scaling law 
(or linear relation in the log-log plot). 
Correspondingly, the local scaling dimension 
$D_m(\epsilon)$ varies as a function of
$\epsilon$. 
At small energy scale, the values of $D_m(\epsilon)$ gradually
increase as the energy scale $\epsilon$ increases, and almost follow
the GOE line. As the energy scale $\epsilon$ increases further,
the values of $D_m(\epsilon)$ bend downwards
and then again turn to increase.
The behavior of $D_m(\epsilon)$
around this energy scale clearly deviates from the GOE limit.
The value of the energy scale where the deviation is maximum is
roughly $\epsilon_{\rm f}\simeq140d\simeq1.7$MeV. From these 
observations, we again find that at small energy scale,
fluctuation of the strength function is essentially governed by the GOE,
as one knows from other measures such as the NND or the strength 
distribution. The scaling analysis by means of the local scaling dimension,
however, can reveal the new feature which may not be realized by other measures.
Namely, the GOE fluctuation is limited to the small energy scale, while,
at intermediate energy scales, a different fluctuation from the GOE
arises.

One may wonder that the statistical error could produce a fictitious dip,
since the number of levels considered 804 is not so large.
Of course, there may be a statistical error. Accordingly, we should take the 
energy scale $\epsilon_{\rm f}\simeq140d\simeq1.7$MeV 
mentioned above as a rough estimate.
The existence of the dip itself, however, is not due to a statistical error. 
This would be verified from a systematic analysis in the next subsection.
Note that the simple estimate of the standard deviation, of 
the $m$th local scaling dimension is 
$\sigma(D_m)\simeq\sqrt{3}m/\sqrt{N_{\rm tot}}$ \cite{aiba}, 
and in the present case
for the second local scaling dimension,  $\sigma(D_2)\simeq0.12$.

A type of the smoothing procedure to obtain the normalized strength function 
does not much affect the results. Indeed, we verified that the Gaussian smoothing 
instead of the Strutinsky smoothing leads to the essentially same results.

\subsection{Origin of the Deviation from the GOE}

We shall investigate the origin of the observed 
energy scale $\epsilon_{\rm f}=1.7$MeV characteristic to 
the fine structures in the isoscalar quadrupole strength function.

\subsubsection{Doorway damping mechanism}

\begin{figure}
\begin{center}
  \begin{minipage}{7.5cm}
       \psfig{file=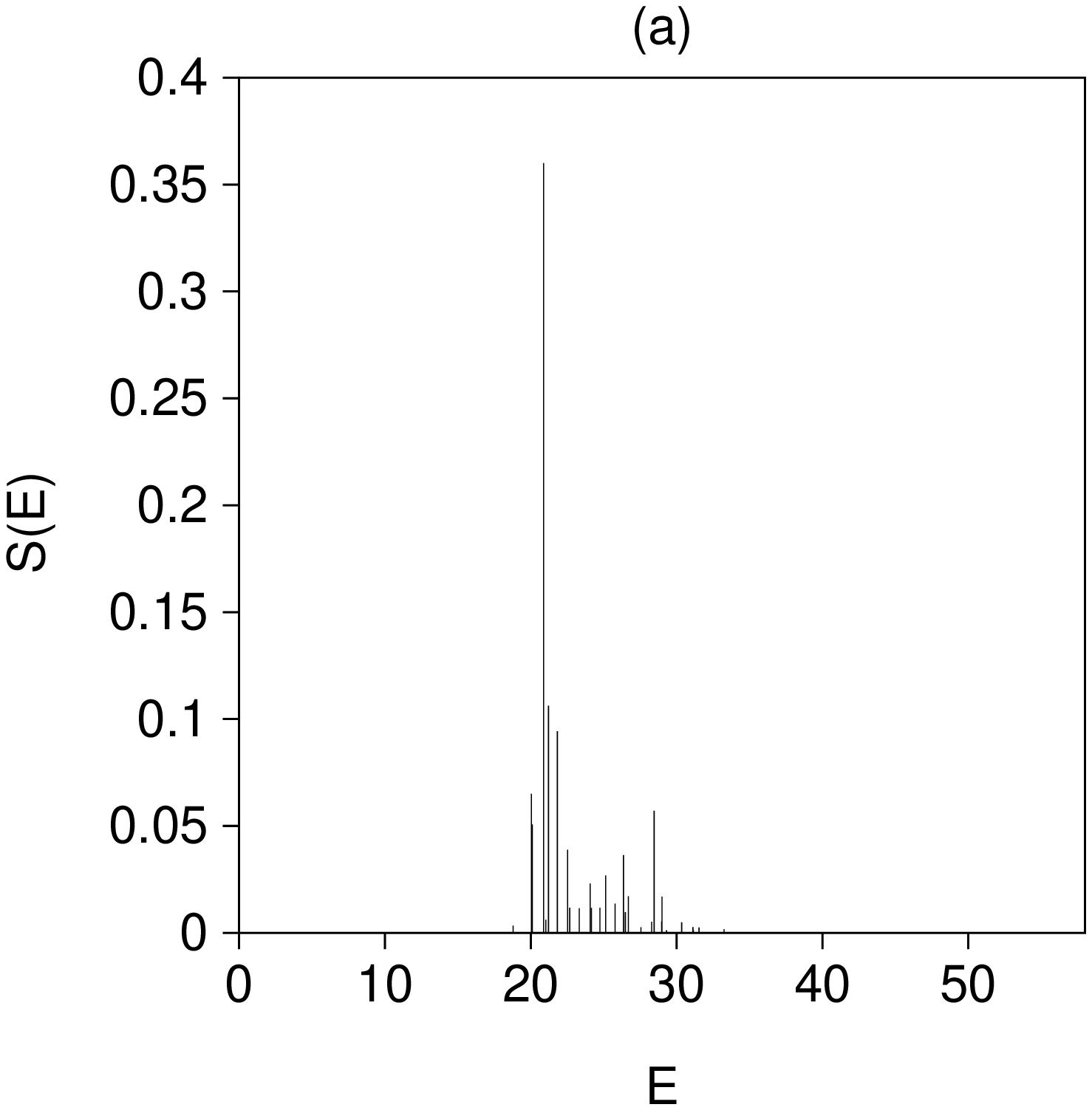,width=7.5cm}
  \end{minipage}
  \begin{minipage}{7.5cm}
       \psfig{file=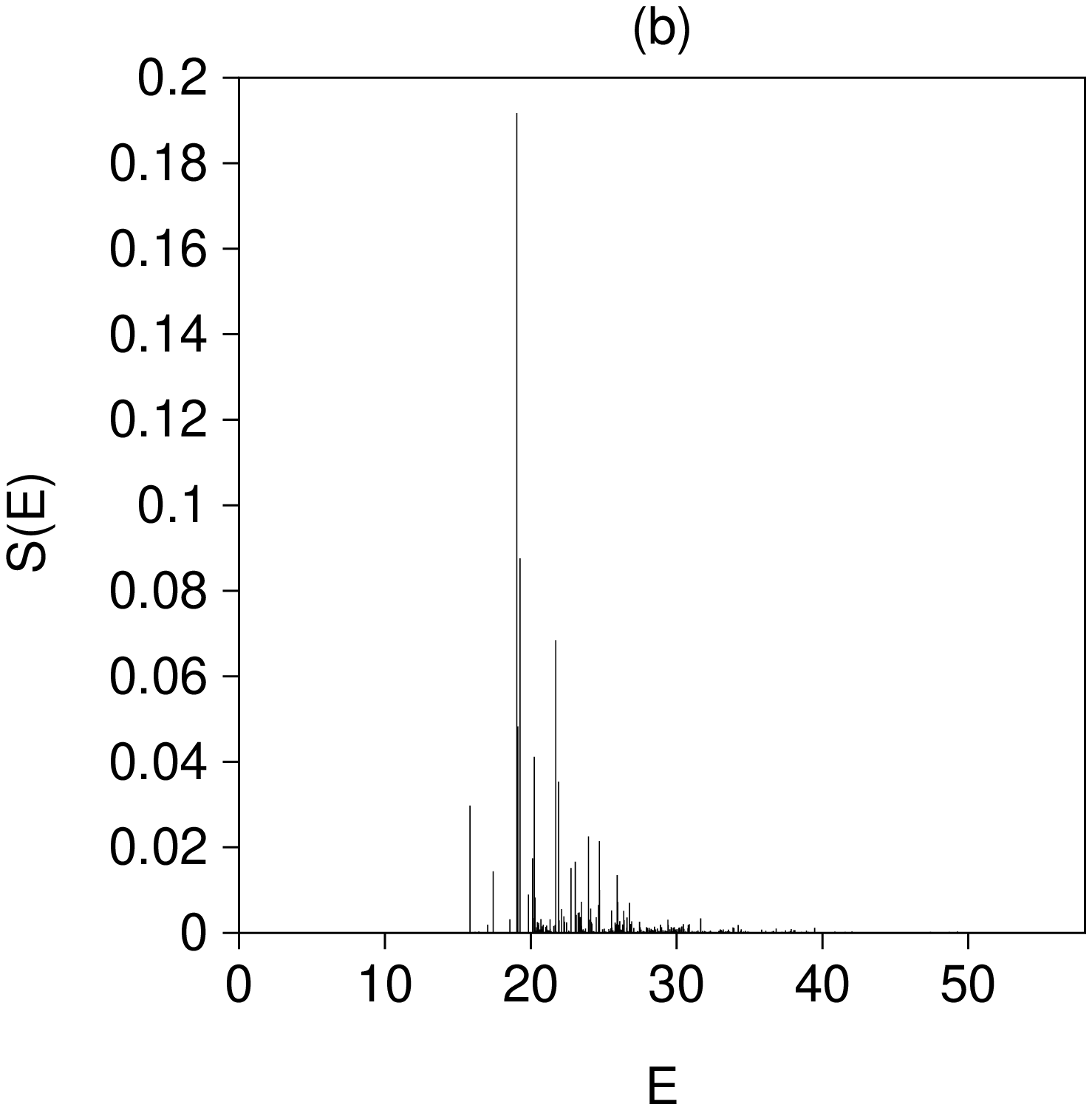,width=7.5cm}
  \end{minipage}
\end{center}
 \caption{
(a) The strength function for the ISGQR obtained within the Tamm-Dancoff
approximation and (b) the strength function for the ISGQR by omitting
the coupling among the unperturbed 2p2h states.
}
\label{fig_tdstrfun}
\end{figure}

For this purpose we consider a damping mechanism by separating
the shell model space to 1p1h and 2p2h subspaces, and we focus on
the residual interactions within and between the subspaces.
The giant resonance is a collective vibrational
state whose dominant component is a coherent superposition
of 1p1h configurations. The collective state
can be described by the Tamm-Dancoff (TD) approximation,
which corresponds to a truncated shell model where only 1p1h
configurations are taken into account. The result of the
TD approximation (TDA) is shown in Fig.\ \ref{fig_tdstrfun}(a), 
and is compared with the
full shell model calculation including up to 2p2h's 
(Fig. \ref{fig_strfun}(a)).
It is seen that 
the peak position is well reproduced by
the TDA. Note also that
the isoscalar quadrupole strength in the TDA
is spread over many eigenstates, which are
correlated 1p1h states.
This spreading of the quadrupole strength among 1p1h states 
is often called the Landau damping.  
Comparing Figs.\ \ref{fig_tdstrfun}(a) and\ \ref{fig_strfun}(a),
 the Landau damping 
is found to give dominant contribution to the total FWHM of
the strength function. 

If we take into account the 2p2h configurations, the
interaction acting between 1p1h 
and 2p2h states (abbreviated as $V_{12}$ ) takes part in. 
Since the 2p2h states have much more degrees of freedom,
$V_{12}$ causes spreading of the TD 1p1h states. 
Because of this spreading of 1p1h states, 
the quadrupole strength is also fragmented by $V_{12}$ on top of the Landau
damping at the TDA level. To see the effect of $V_{12}$, we performed
a calculation where unperturbed 2p2h states and the interaction
$V_{12}$ is introduced. 
The resultant strength distribution, shown in Fig.\ \ref{fig_tdstrfun}(b),
 appears
similar to the final result (Fig.\ \ref{fig_strfun}(a)) 
as far as the gross profile
of the strength distribution is concerned. 

Note also that there exists
the residual interaction acting among 2p2h states, which we call 
$V_{22}$ in the following.  $V_{22}$ causes configuration mixing among
the 2p2h states.  Comparison between Figs.\ \ref{fig_tdstrfun}(b)
 and \ref{fig_strfun}(a) indicates
effects of $V_{22}$ on fine details of strength distribution, although
the gross profile of strength distribution is not influenced very much. 

The above observations lead us to the following doorway damping picture.
The quadrupole strength distribution is fragmented
first in the 1p1h subspace. The 1p1h TD states
then spread further over 2p2h states through $V_{12}$. Here the 1p1h TD states
may be considered as doorway states in the whole damping processes.
The 2p2h states on the other hand
may be regarded as states that play roles of background
in the main damping processes since
they do not dominate the total damping width, but they
influence the fine structures through the coupling to 
the doorway states. 

\begin{figure}
\begin{center}
 \begin{minipage}{7.5cm}
       \psfig{file=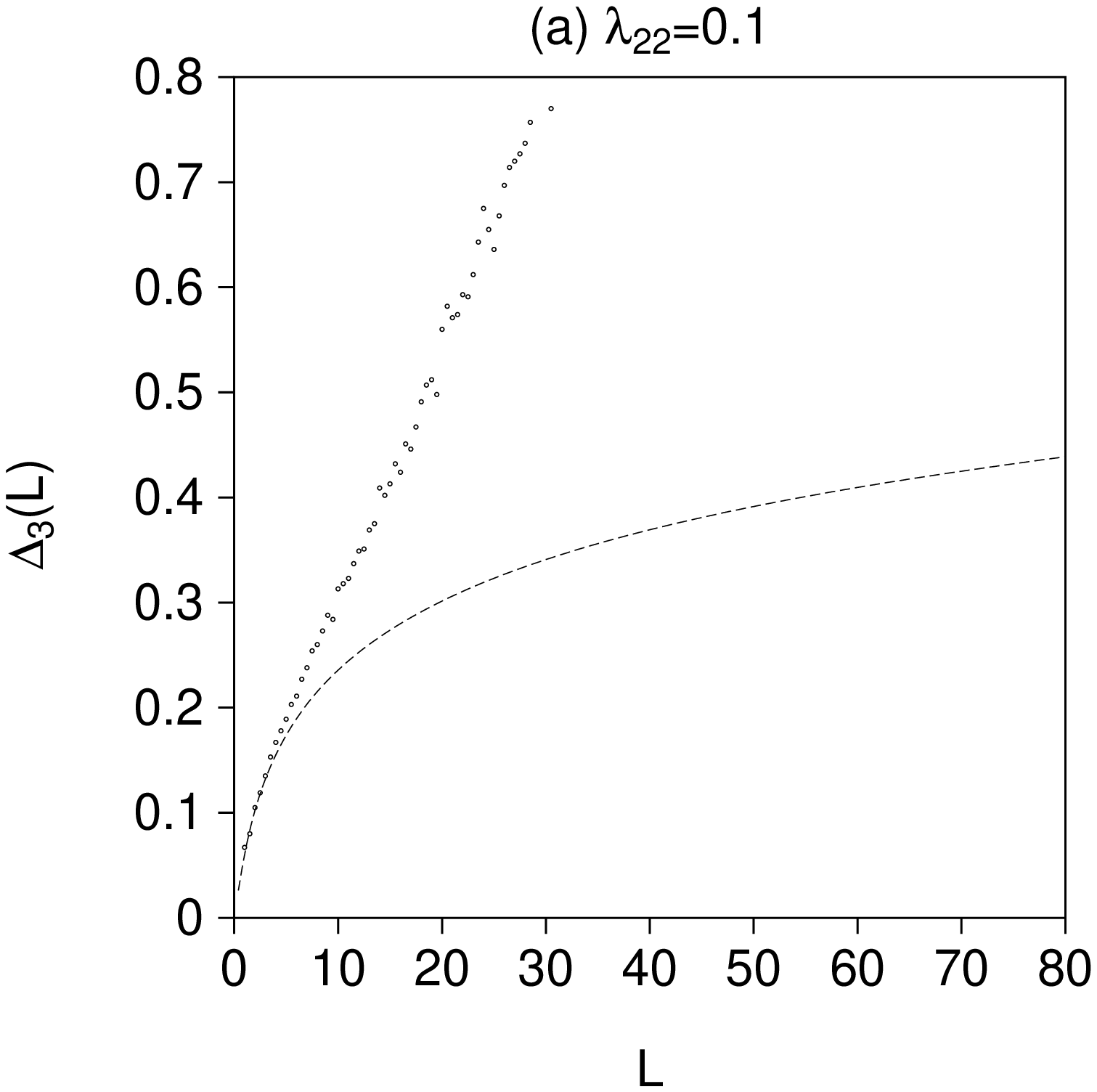,width=7.5cm}
 \end{minipage}
 \begin{minipage}{7.5cm}
       \psfig{file=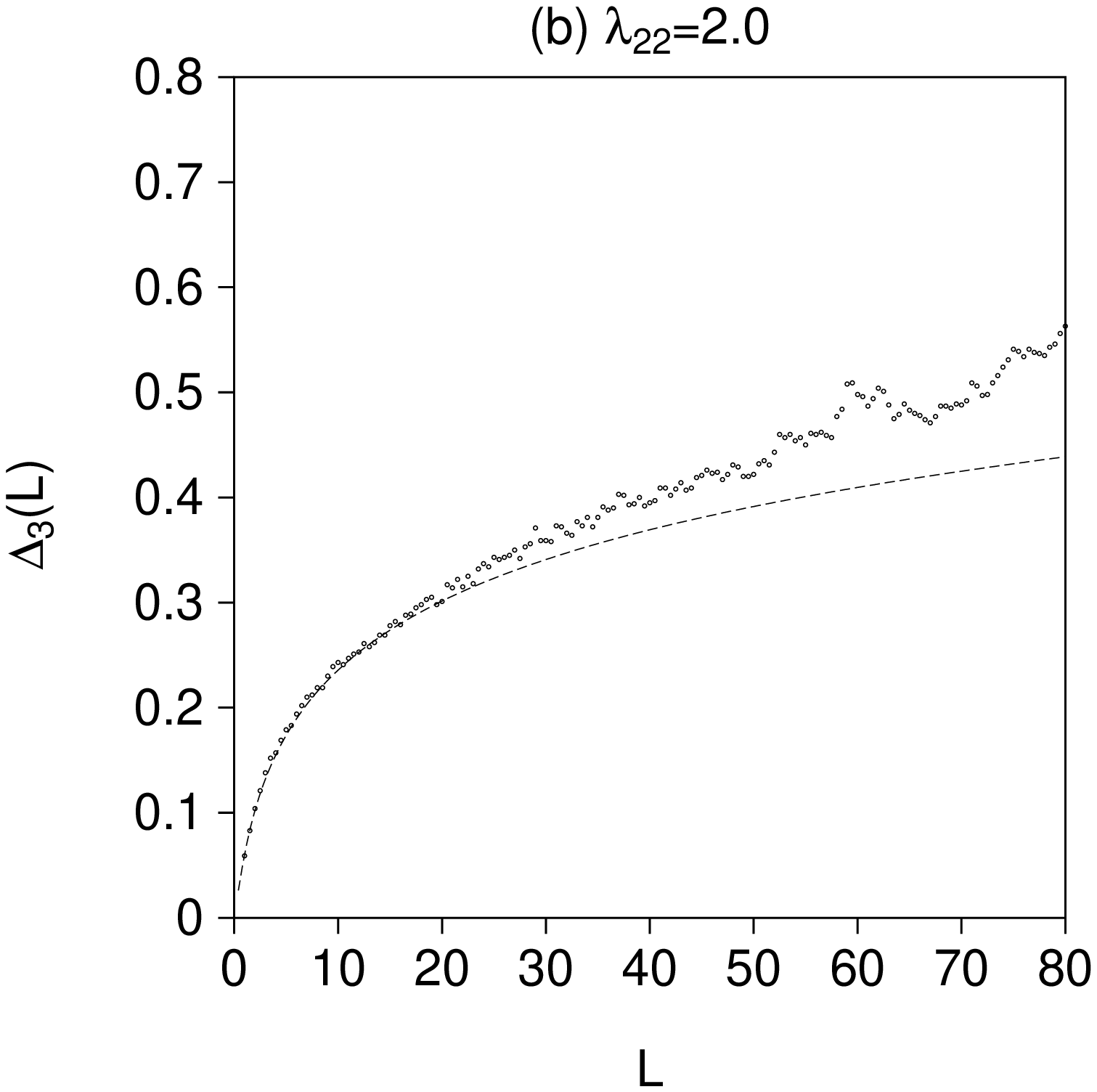,width=7.5cm}
 \end{minipage}
\end{center}
 \caption{
The $\Delta_3$ statistics with rescaling the interaction strength
$V_{22}$ by a factor $\lambda_{22}$, for (a)$\lambda_{22}=0.1$ and (b)$\lambda_{22}=2.0$.
The dashed curve represents the $\Delta_3$ for the GOE level fluctuation.
We can find the large dependence of the  $\Delta_3$ on the value of
$V_{22}$. We also verified that the $\Delta_3$ is insensitive to the
change of the value of $V_{12}$, namely the behavior of $\Delta_3$ at
$\lambda_{12}=0.1$ or $\lambda_{12}=2.0$ is almost same as that at $\lambda_{12}=1.0$,
where $\lambda_{12}$ represents the rescaling factor of the interaction strength
$V_{12}$.
}
\label{fig_delta3_V22}
\end{figure}

Assuming the above picture, we can consider energy scales
that may be relevant to the fine structures of the quadrupole strength distribution.
Concerning the 1p1h TD states  which are considered as the doorway states, 
we have as energy scales
1) the spreading width (which we denote $\gamma_{12}$) 
of the TD 1p1h states caused by the interaction $V_{12}$,
in addition to 
2) the average level spacing $d_{1p1h}$ between the 1p1h states.
Concerning the 2p2h states,
3) the level spacing $d_{2p2h}$ between 2p2h states is to be noted.
($d_{2p2h}$ is almost identical to the level spacing $d$ 
for the whole set of energy spectra since the number of 2p2h
states is much larger than that of 1p1h states. )
Note also that mixing
among 2p2h states caused by the interaction $V_{22}$ is characterized
by 3) the spreading width ($\gamma_{22}$ named in the following) 
of 2p2h states. 
We have found previously that, in the case of a schematic model which
incorporate doorway states coupled to background states, 
the spreading width of doorway states
(corresponding to $\gamma_{12}$ in the present context) determines
the energy scale where the local scaling dimension deviates from 
the generic GOE limit \cite{aiba}.
Note also that the deviation from the GOE is seen in the energy 
level fluctuation by the analysis of the $\Delta_3$ statistics. 
The energy level fluctuation is sensitive to the residual interaction 
that causes mixing among unperturbed configurations.
We found that the energy level fluctuation is  governed by 
$V_{22}$, but not by $V_{12}$, as shown in Fig.\ \ref{fig_delta3_V22}. 
Thus we can consider another possibility that the energy scale of the 
deviation from the GOE limit may be related to the spreading
width $\gamma_{22}$ of the 2p2h states.
In the following, we shall focus
on the effects of $V_{12}$ and $V_{22}$ 
which are connected to the two spreading widths 
$\gamma_{12}$ and $\gamma_{22}$.

\subsubsection{Spreading widths of 1p1h and 2p2h states}

Let us quantify $\gamma_{12}$ and $\gamma_{22}$.
If we assume the Fermi golden rule, we obtain
a simple estimate of $\gamma_{12}$ by
$
\gamma_{12}^{\rm FG} = 2\pi \overline{\bra{1p1h}V_{12}\ket{2p2h}}^2 /d_{2p2h}
$
where $\overline{\bra{1p1h}V_{12}\ket{2p2h}}$
is a root mean square value of the matrix elements between
1p1h and 2p2h states.
Similarly, we can 
make an estimation of the spreading width $\gamma_{22}$ as 
$
\gamma_{22}^{\rm FG} = 2\pi \overline{\bra{2p2h}V_{22}\ket{2p2h}}^2 /d_{2p2h}.
$
The above estimate gives 
$\gamma_{12}^{\rm FG}\simeq5.3 $MeV, and 
$\gamma_{22}^{\rm FG}\simeq6.0 $MeV. Here we have
inserted $\overline{\bra{1p1h}V_{12}\ket{2p2h}}=98$ keV, 
$\overline{\bra{2p2h}V_{22}\ket{2p2h}}=104$ keV, 
$d_{2p2h}=11.3$keV, which are evaluated
for all the unperturbed 1p1h and 2p2h states in the
energy interval of $E=20-30$MeV. Note also $d_{1p1h}\approx 0.5$MeV
for the 1p1h states in the same interval. Very similar values
are obtained for $\overline{\bra{1p1h}V_{12}\ket{2p2h}}$ when we use the
correlated 1p1h states (TD states) instead of unperturbed 1p1h's.

\begin{figure}
\begin{center}
 \begin{minipage}{7.5cm}
       \psfig{file=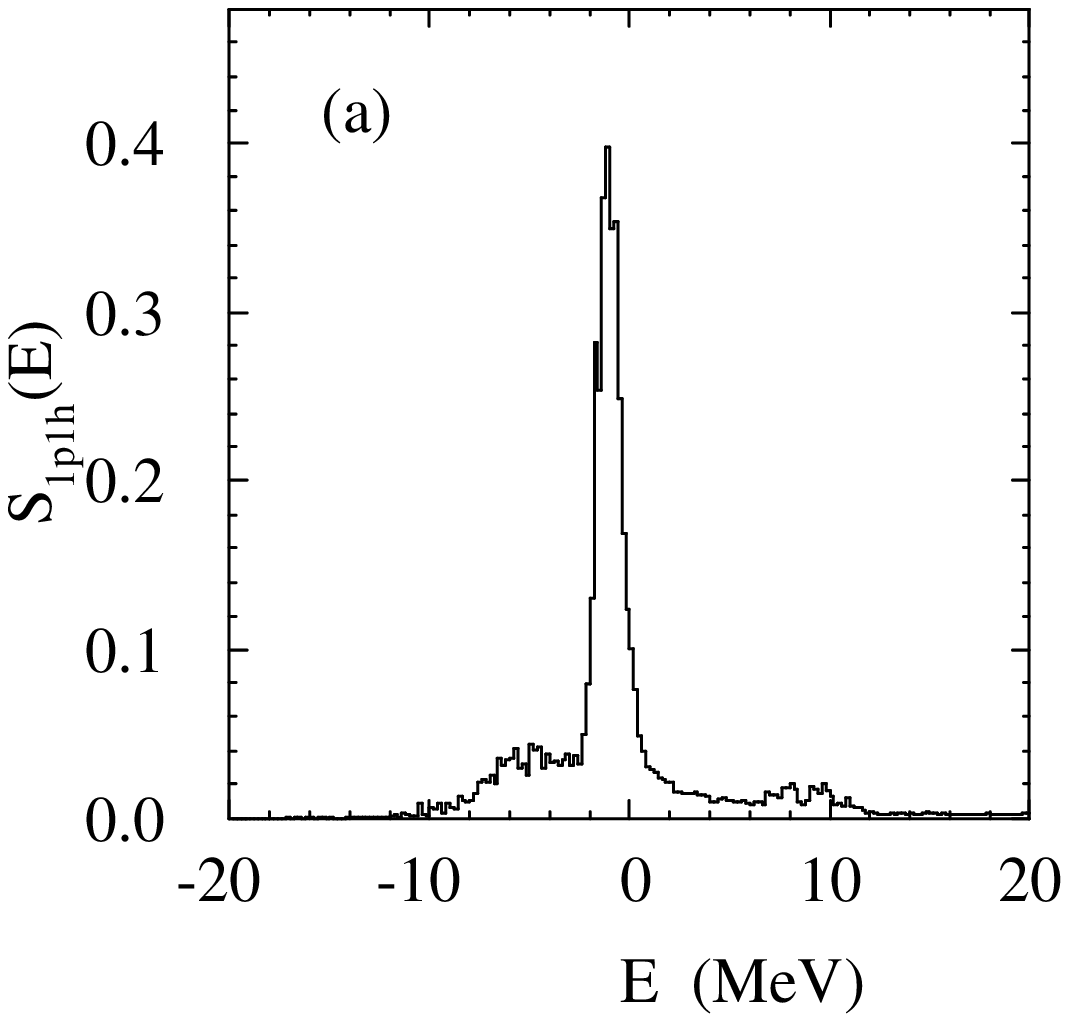,width=7.5cm}
 \end{minipage}
 \begin{minipage}{7.5cm}
       \psfig{file=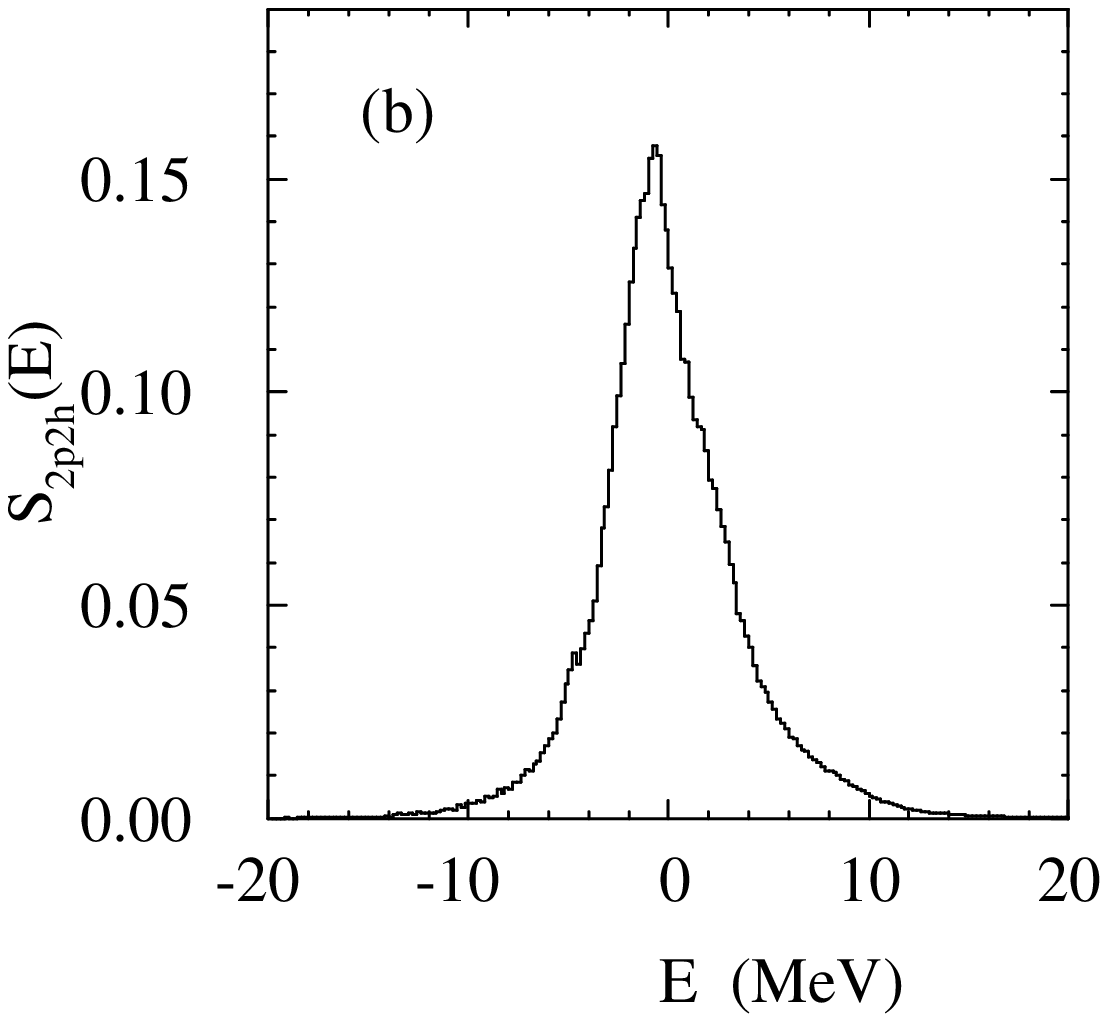,width=7.5cm}
 \end{minipage}
\end{center}
 \caption{
(a) The strength function of the 1p1h state averaged over all 1p1h states
within the energy interval $E=20-30$MeV and (b) the strength function of 
the 2p2h states averaged over the same energy interval.
}
\label{fig_1p1hstrfun}
\end{figure}

It is also possible to evaluate $\gamma_{12}$ and $\gamma_{22}$ in
a more direct way by using the strength functions of 1p1h
and 2p2h states. Strength function of a 1p1h state 
$\ket{i,1p1h}$ is given
by 
\begin{equation}
S_{i,1p1h}(e)=\sum_{a}\left<i,1p1h|a\right>^2 \delta(e-(E_a-E_{i,1p1h})),
\label{eq_1p1hstrfun}
\end{equation}
where $\ket{a}$ and $E_a$ are the eigenstates of the full calculation
and their energies, respectively.  $e$ denotes 
the relative energy from the energy centroid 
$E_{i,1p1h}$ of the 1p1h state considered. 
Evaluating an average of $S_{i,1p1h}(e)$ over all 
the 1p1h states, the average strength function of
1p1h states is obtained, as shown in Fig.\ \ref{fig_1p1hstrfun}(a). Its FWHM 
gives an evaluation of the average spreading width $\gamma_{12}$
of 1p1h states. Similar method can be used to evaluate
the spreading width $\gamma_{22}$ of the 2p2h states
(cf. Fig.\ \ref{fig_1p1hstrfun}(b)).
This direct method gives $\gamma_{12}^{\rm dm}=1.5$MeV 
and $\gamma_{22}^{\rm dm}=5.2$MeV.

It is noted that the values of the spreading width $\gamma_{12}$ are very
different between the golden rule and the direct method.
To investigate the difference,
we evaluated root mean square values of interaction matrix elements of
$V_{12}$ for individual
1p1h states, and found that 
about one third of the 1p1h states have significantly
large value $\simeq 120-200$ keV as compared with the other 1p1h states
which have small value  $\simeq 30-70$keV. Corresponding golden
rule estimates of the spreading width are $\gamma_{12}^{\rm FG}=8.3-22$MeV and
$\gamma_{12}^{\rm FG}=0.5-2.8$MeV, respectively. 
The FWHM of the average strength function of 1p1h states
used in the direct method is dominated by the 
1p1h states having small FWHM. Accordingly
the direct method gives much smaller value ($\gamma_{12}^{\rm dm}=1.5$ MeV)
than that of the golden rule ($\gamma_{12}^{\rm FG}=5.3$ MeV). 

\subsubsection{Dependence of LSD on $V_{12}$ and $V_{22}$}

In order to identify the origin of the
observed energy scale $\epsilon_{\rm f}=1.7$MeV,
we shall analyze the dependence of the fine structures of the strength functions
on $V_{12}$ and $V_{22}$, which 
are intimately related to the spreading widths $\gamma_{12}$ and
$\gamma_{22}$. 
For this analysis we rescale the interaction strength $V_{12}$ and
$V_{22}$ by factors $\lambda_{12}$ and $\lambda_{22}$ respectively, and 
look into the local scaling dimension to see how the fine structures change. 
 
\begin{figure}
\begin{center}
  \begin{minipage}{7.5cm}
       \psfig{file=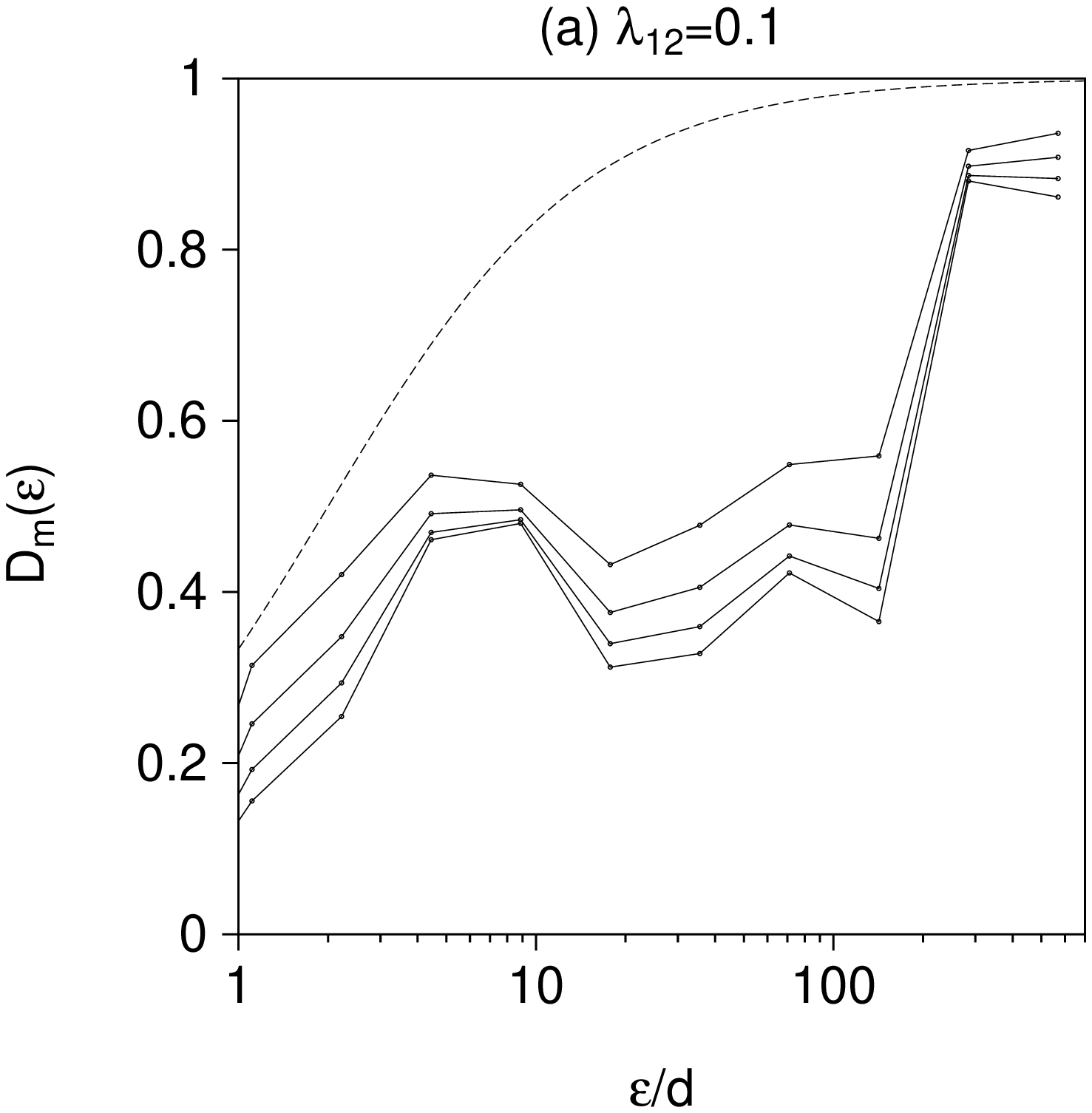,width=7.5cm}
  \end{minipage}
  \begin{minipage}{7.5cm}
       \psfig{file=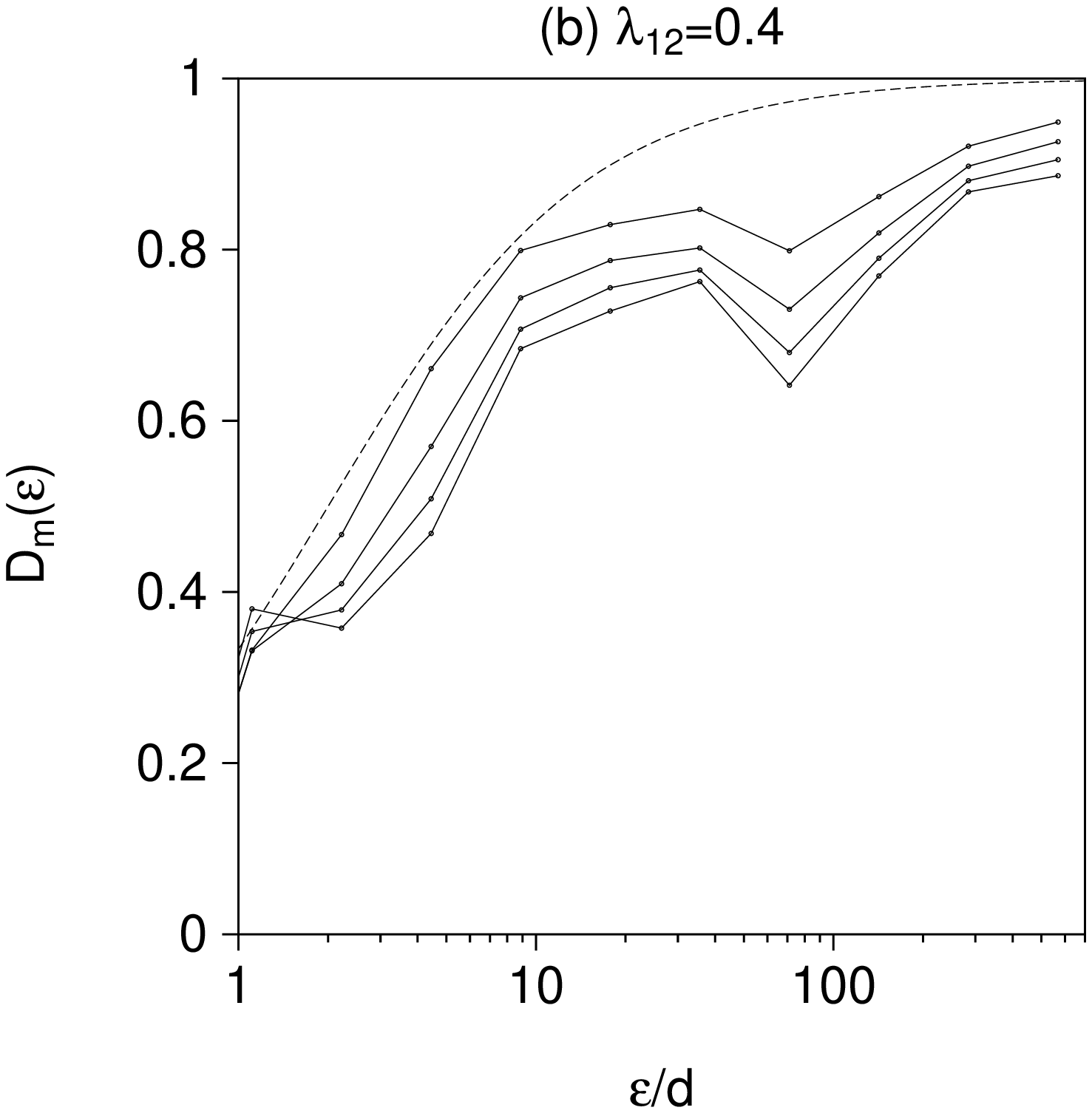,width=7.5cm}
  \end{minipage}
\end{center}
\begin{center}
  \begin{minipage}{7.5cm}
       \psfig{file=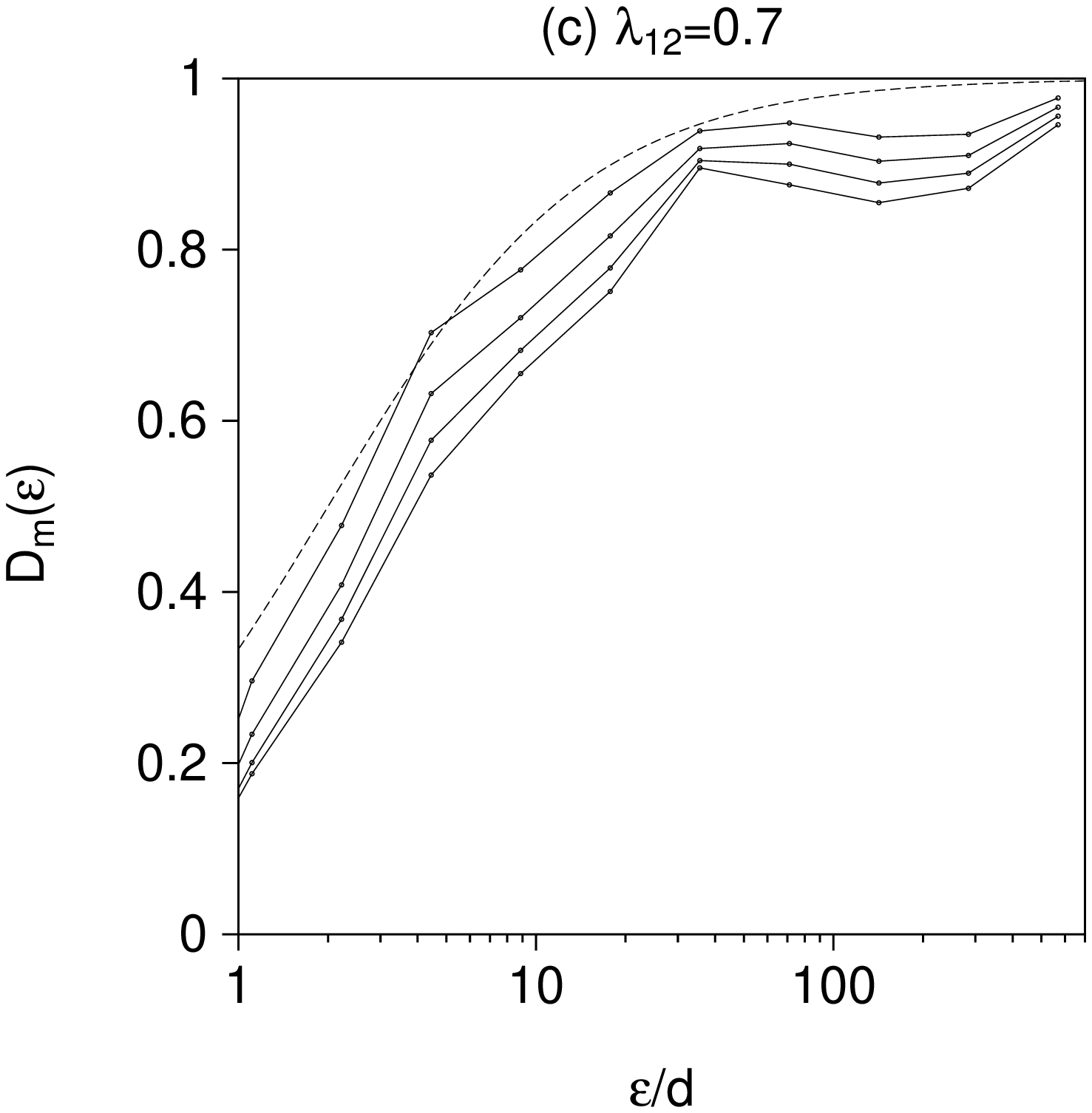,width=7.5cm}
  \end{minipage}
  \begin{minipage}{7.5cm}
       \psfig{file=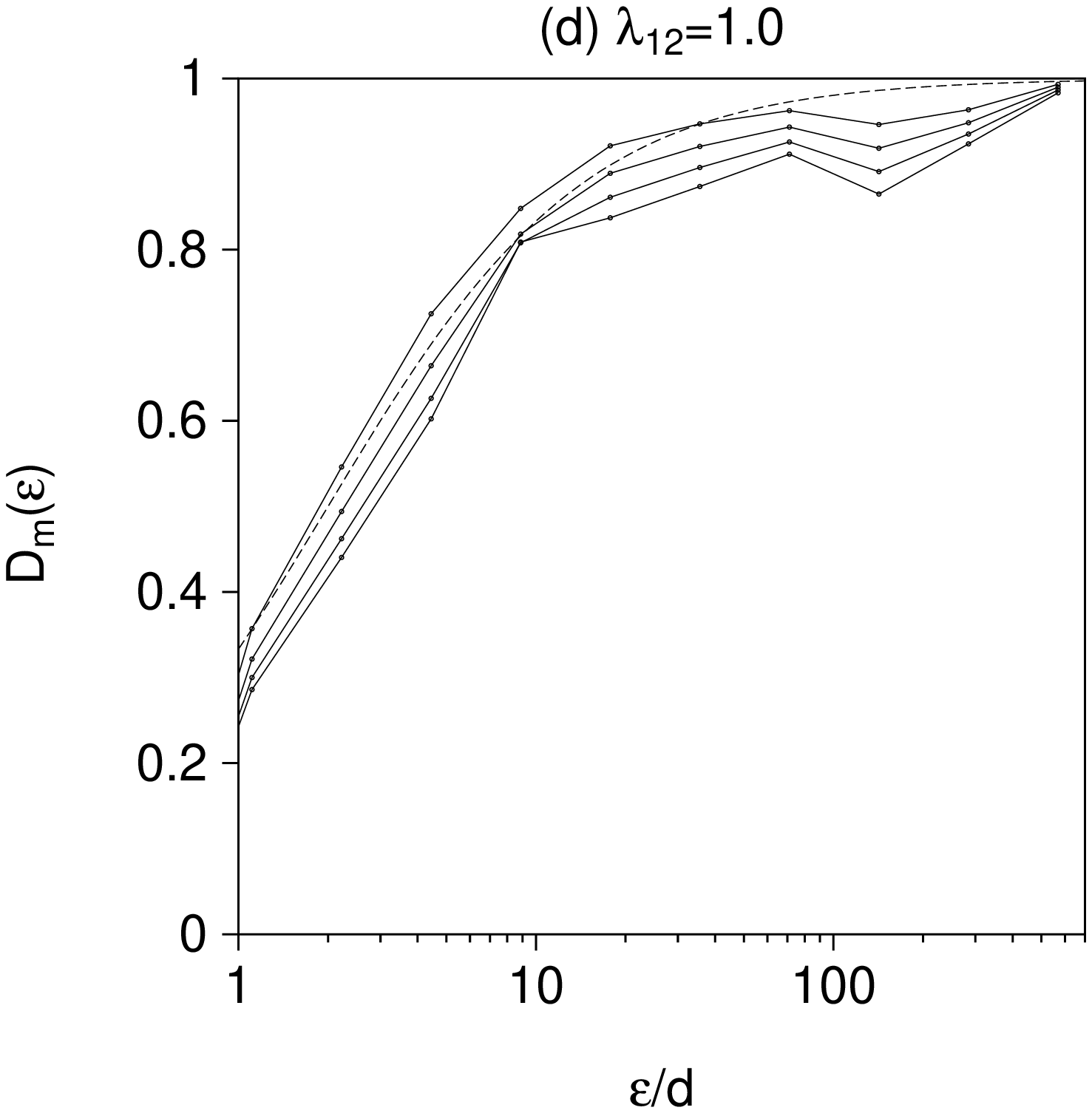,width=7.5cm}
  \end{minipage}
\end{center}
\begin{center}
  \begin{minipage}{7.5cm}
       \psfig{file=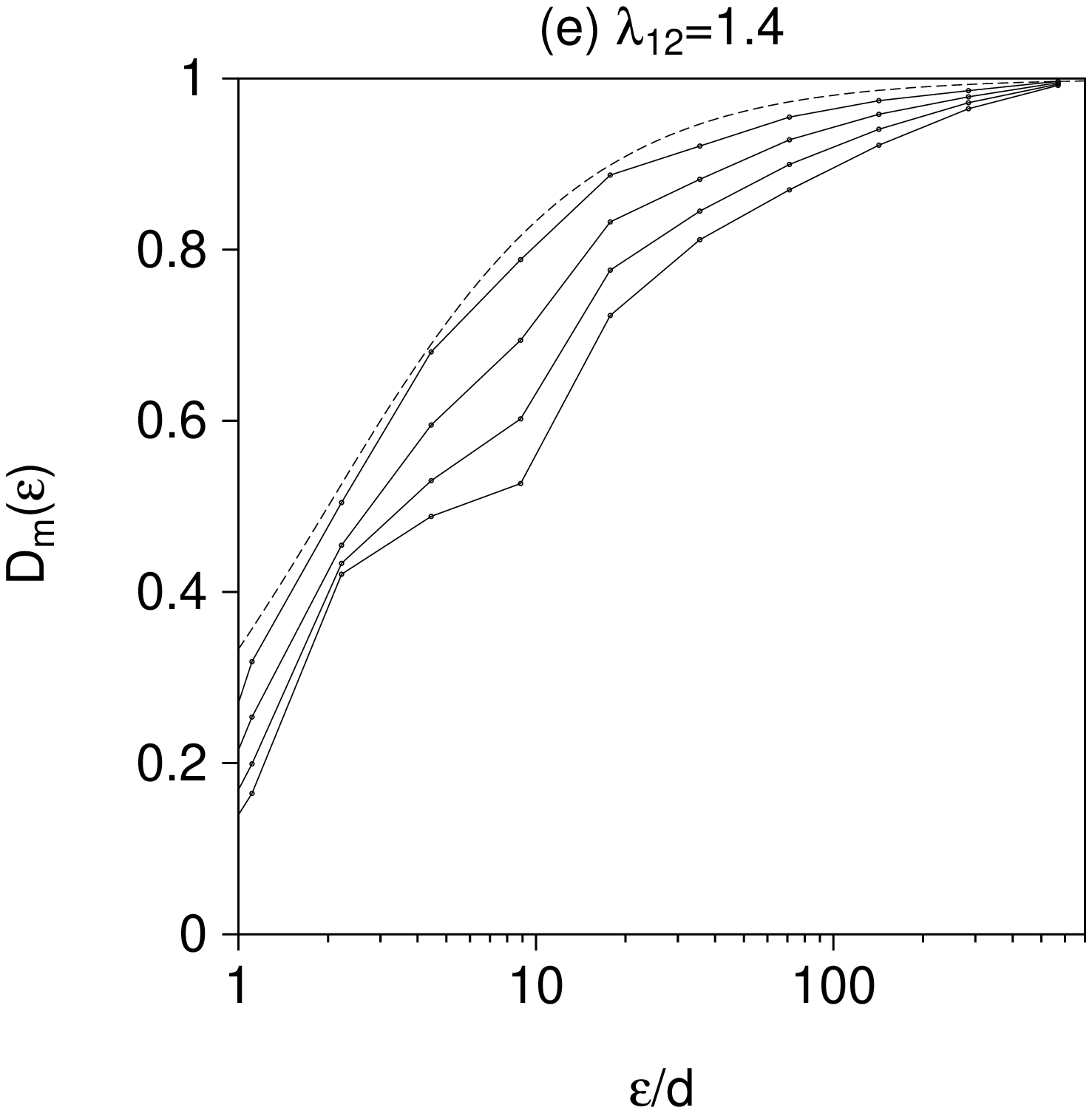,width=7.5cm}
  \end{minipage}
  \begin{minipage}{7.5cm}
       \psfig{file=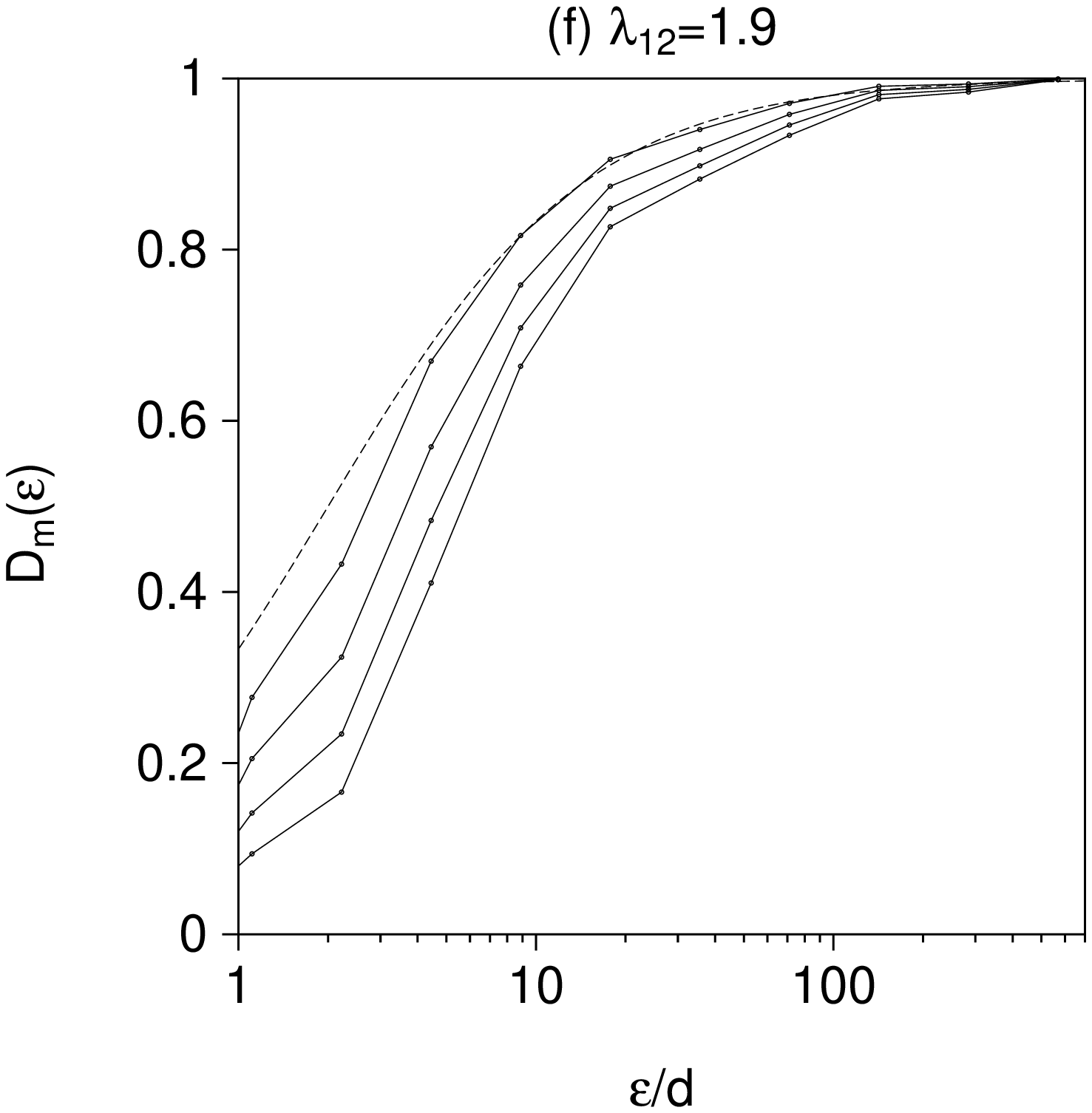,width=7.5cm}
  \end{minipage}
\end{center}
 \caption{
The local scaling dimensions $D_m(\epsilon)$ ($m=2$ to 5) for the ISGQR calculated
by changing values of $\lambda_{12}$, which is chosen as (a) $\lambda_{12}=0.1$,
(b)$\lambda_{12}=0.4$, (c)$\lambda_{12}=0.7$, (d)$\lambda_{12}=1.0$, (e)$\lambda_{12}=1.4$, and 
(f)$\lambda_{12}=1.9$.
Here $\lambda_{12}$ represents the ratio between the adopted values of the Hamiltonian 
matrix elements between the TD 1p1h 
states and 2p2h states, and those of original values.
The dashed curve represents $D_2(\epsilon)$ for the GOE.
}
\label{fig_V12}
\end{figure}

It is seen from Fig.\ \ref{fig_V12}  
that the fine structures probed by the local scaling dimension $D_m(\epsilon)$ 
strongly depend on $\lambda_{12}$.
When the value of $\lambda_{12}$ is small, $\lambda_{12}=0.1$,
the LSD exhibits very strong deviation from the GOE limit 
in almost all energy scales except
at very small energy scale  $\epsilon/d \sim 1$.
As the value of $\lambda_{12}$ increases, the deviation becomes smaller and
the LSD at small energy scales coincides with the GOE limit.
The energy scale $\epsilon^{*}$
where the LSD start to deviate from the GOE limit
moves to larger values as $\lambda_{12}$ increases. 
If $\lambda_{12}$ is increased more than the original value (e.g. $\lambda_{12}=1.7$),
the LSD almost reaches the GOE limit over almost whole energy scales.
The above dependence on $\lambda_{12}$ agrees very well with
a similar behavior found in the simple doorway damping model \cite{aiba}.
(In that schematic model
one can account for the spreading width $\gamma_{\rm dw}$ of the doorway states
analytically.)  We found in the schematic model that the energy scale
$\epsilon^*$ 
is proportional to $\gamma_{\rm dw}$ as $\epsilon^* \approx \gamma_{\rm dw}/5$
and the energy scale $\epsilon^{**}$ where the LSD decreases most steeply 
corresponds to $\gamma_{\rm dw}$. $\epsilon^*$ and $\epsilon^{**}$ moves
with the interaction $v$ between the doorway states and the background
states since $\gamma_{\rm dw}\propto v^2$. This behavior is consistent
with the $\lambda_{12}$ dependence in the present model.
For quantitative evaluation, we compare the spreading
width $\gamma_{12}$ of the 1p1h TD states,
which is considered here as the doorway states,  and 
the energy scale $\epsilon_{\rm f}$
where the deviation of the LSD from the GOE limit is maximum.
Here we adopt the energy scale $\epsilon_{\rm f}$
in place of $\epsilon^{**}$, whose extraction is not easy in
the present model due to fluctuation in the LSD. 
The energy scale $\epsilon_{\rm f}=1.7$MeV found for the original interaction 
(i.e. $\lambda_{12}=1$) agrees with the spreading width $\gamma_{12}$ of 
the 1p1h states if we adopt the value $\gamma_{12}^{\rm dm}=1.5$MeV obtained 
by the direct evaluation.
The agreement is a little
worse if we refer to the value $\gamma_{12}^{\rm FG}=5.3$MeV derived from 
the golden rule estimate. We expect that the fine structures 
are dominated by the group of 1p1h states having
smaller spreading width $\gamma_{12}\sim 1$MeV, while the remaining 1p1h 
states with  $\gamma_{12}>10$MeV, which is larger than the total width 
of the giant resonance, do not contribute to the fine structure. 
The comparison with the spreading width $\gamma_{12}^{\rm dm}$ by
the direct evaluation would be more appropriate since 
$\gamma_{12}^{\rm dm}$ reflects mainly the
first group. (On the other hand, $\gamma_{12}^{\rm FG}$ based on the root 
mean square of interaction matrix elements puts emphasis on the other 
group. See the previous subsection.)
The $\lambda_{12}$-dependence and the comparison with $\gamma_{12}$ indicate
an approximate relation $\epsilon_{\rm f} \sim \gamma_{12}$, for which 
we estimate an ambiguity by a factor of two.

\begin{figure}
\begin{center}
  \begin{minipage}{7.5cm}
       \psfig{file=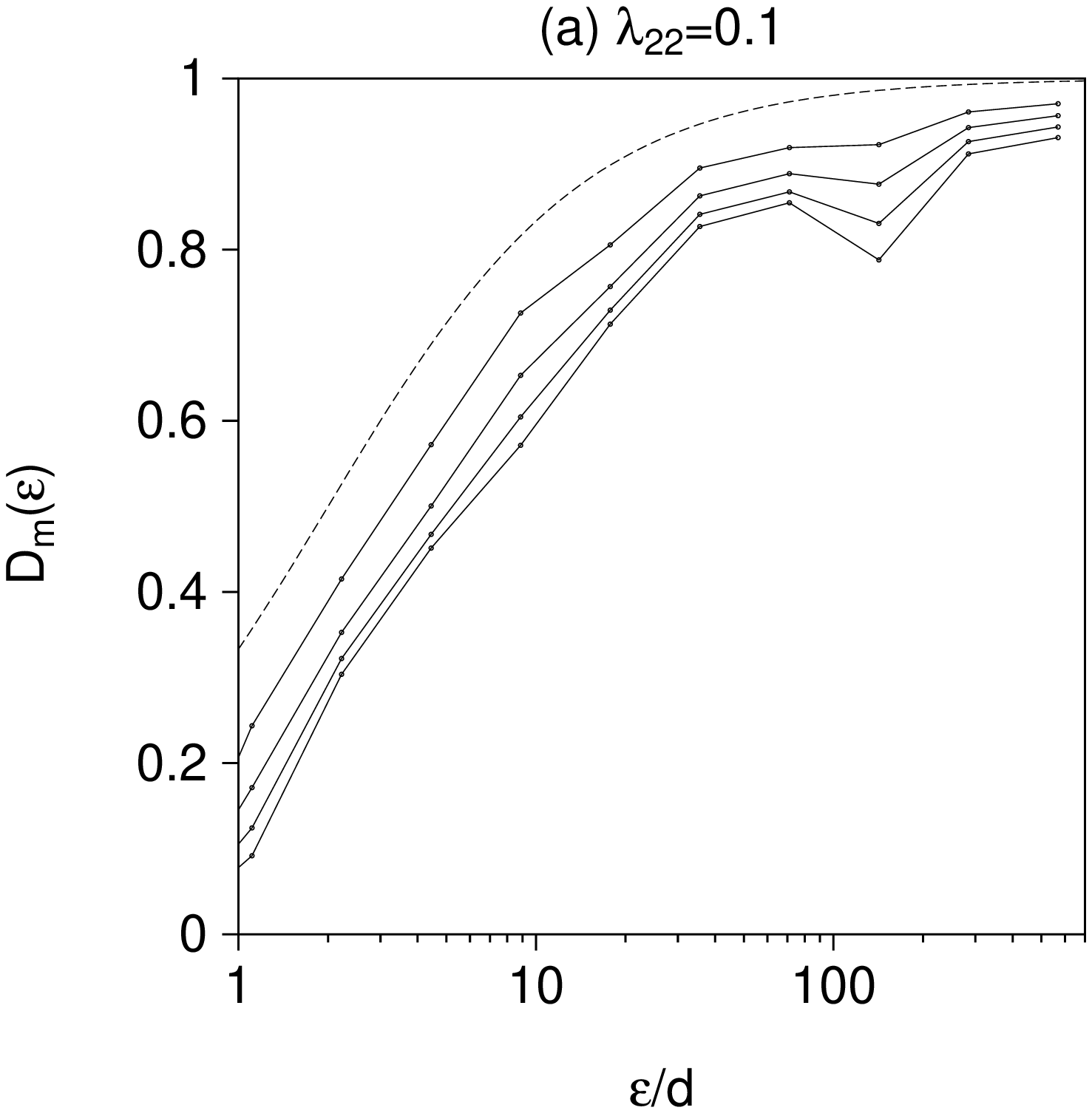,width=7.5cm}
  \end{minipage}
  \begin{minipage}{7.5cm}
       \psfig{file=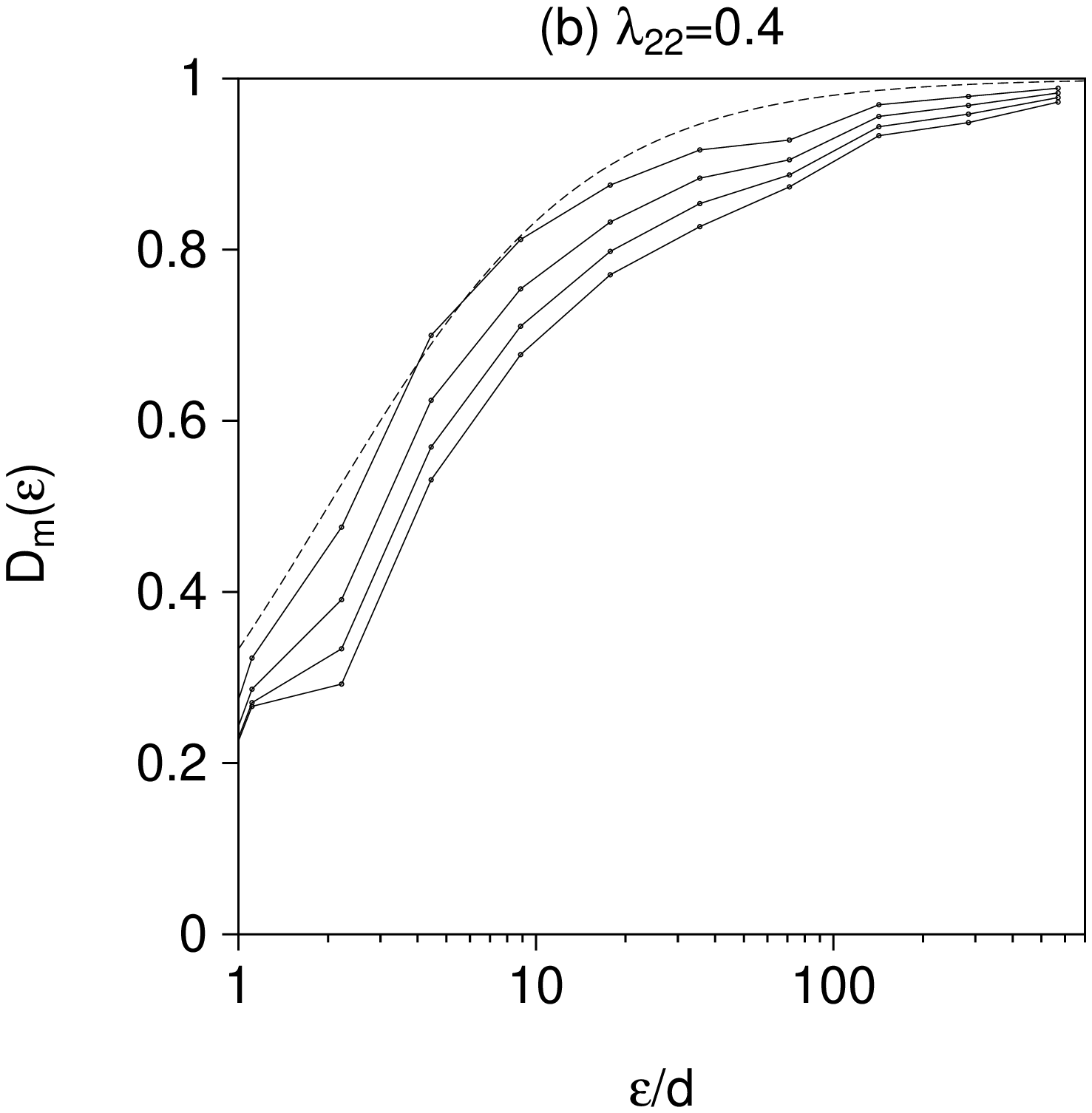,width=7.5cm}
  \end{minipage}
\end{center}
\begin{center}
  \begin{minipage}{7.5cm}
       \psfig{file=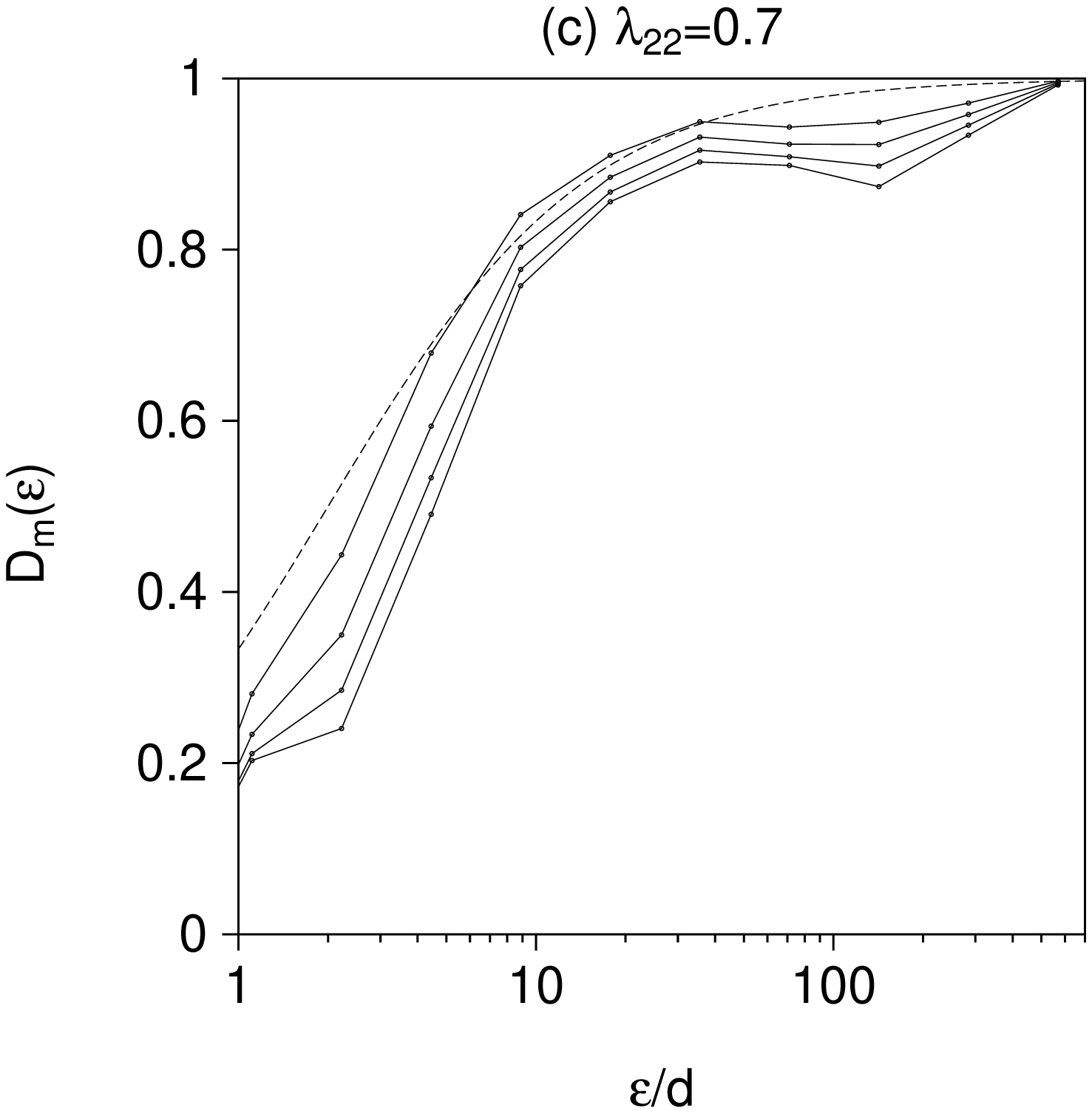,width=7.5cm}
  \end{minipage}
  \begin{minipage}{7.5cm}
       \psfig{file=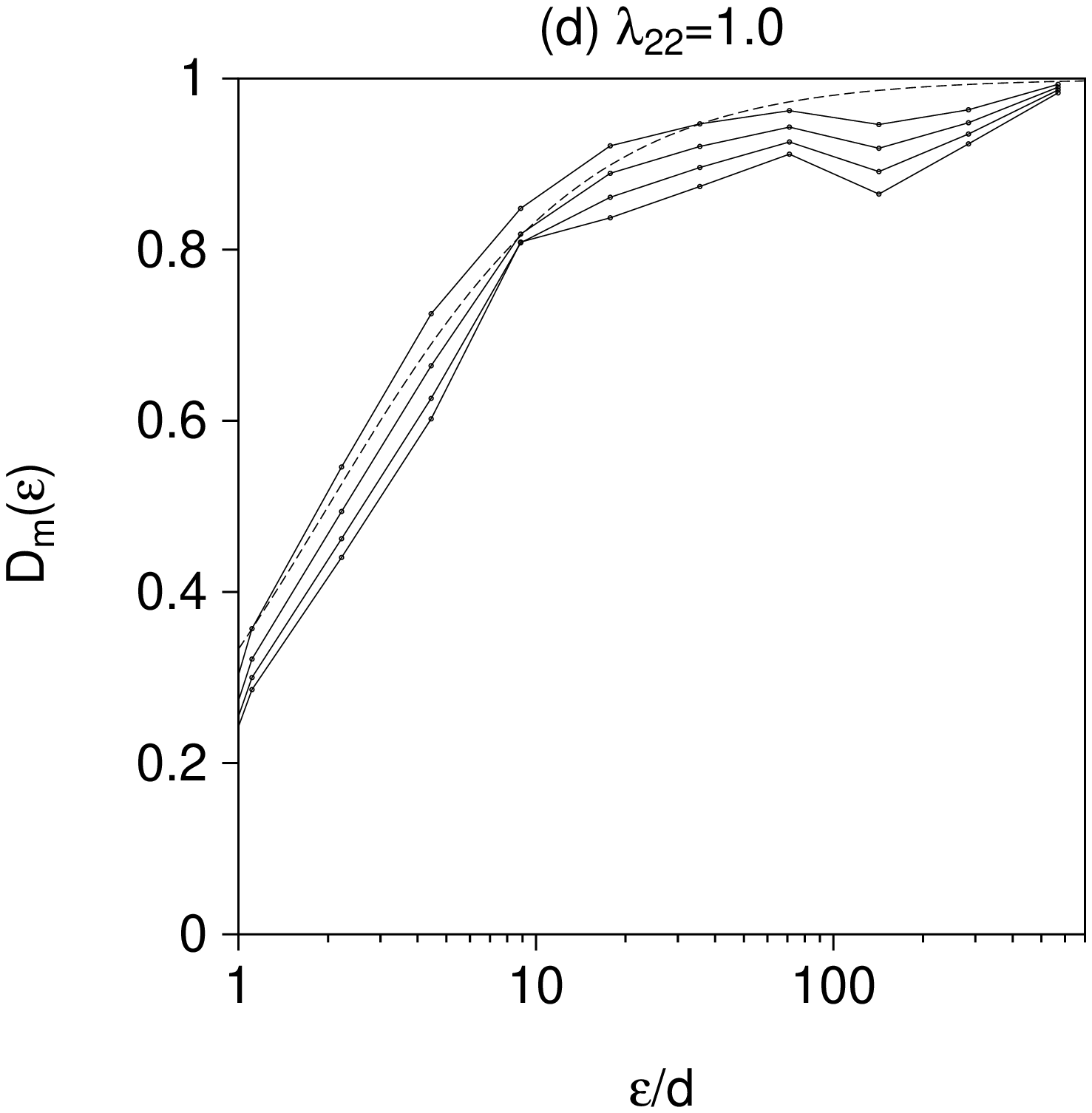,width=7.5cm}
  \end{minipage}
\end{center}
\begin{center}
  \begin{minipage}{7.5cm}
       \psfig{file=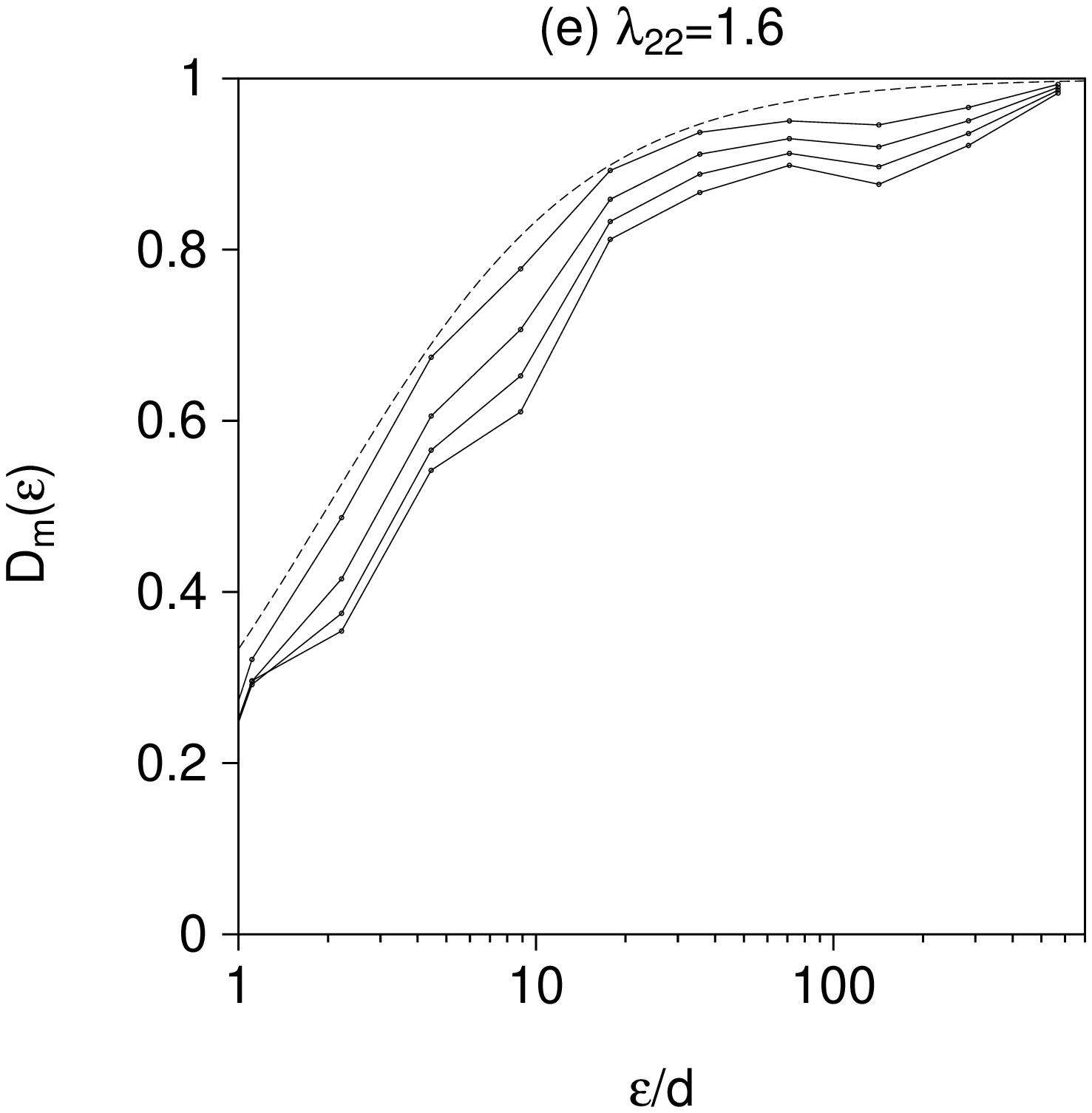,width=7.5cm}
  \end{minipage}
  \begin{minipage}{7.5cm}
       \psfig{file=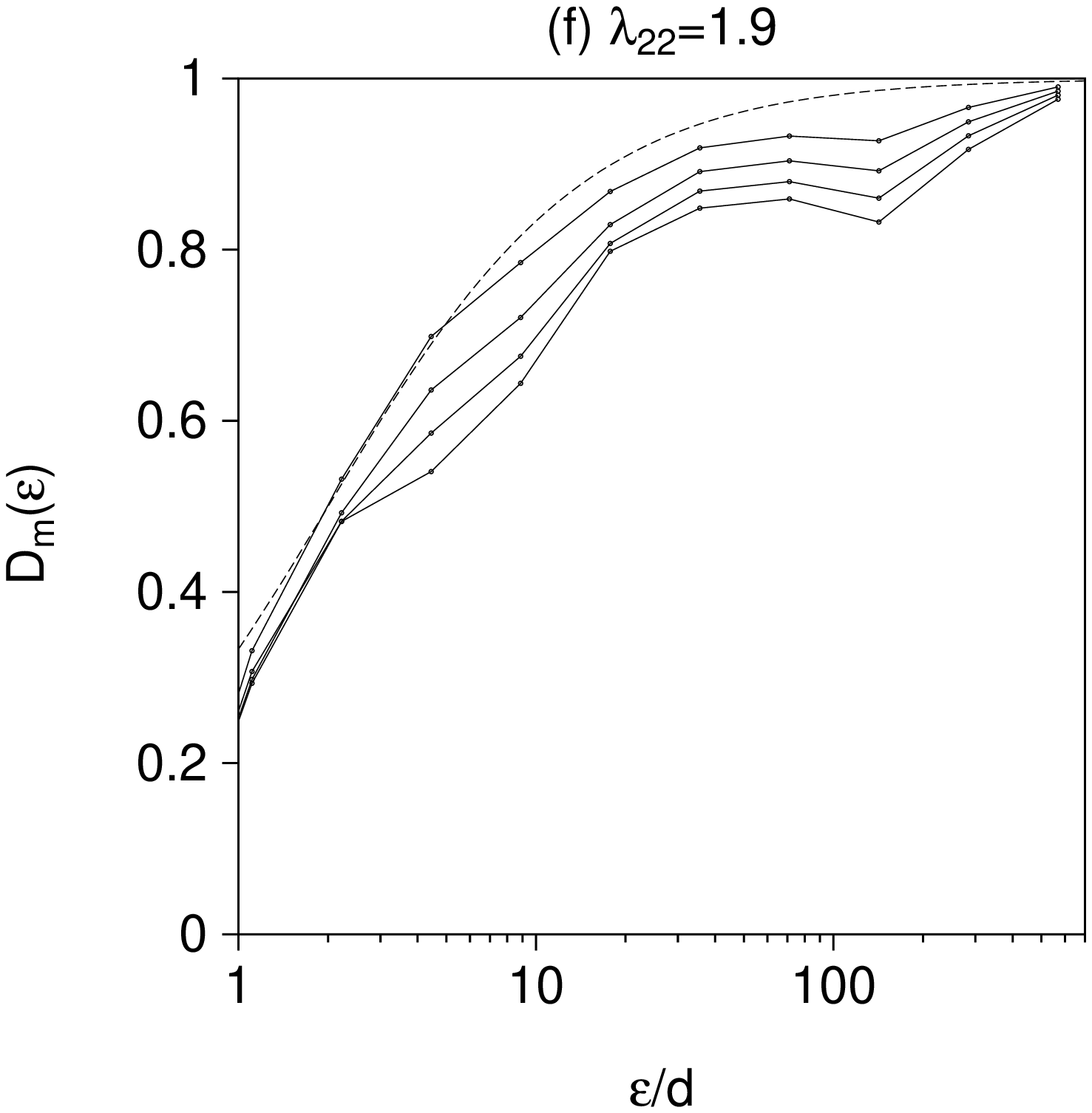,width=7.5cm}
  \end{minipage}
\end{center}
 \caption{
The local scaling dimensions $D_m(\epsilon)$ ($m=2$ to 5) for the ISGQR calculated
by changing values of $\lambda_{22}$, which is chosen as (a) $\lambda_{22}=0.1$,
(b)$\lambda_{22}=0.4$, (c)$\lambda_{22}=0.7$, (d)$\lambda_{22}=1.0$, (e)$\lambda_{22}=1.6$, and 
(f)$\lambda_{22}=1.9$.
Here $\lambda_{22}$ represents the ratio between the adopted values of the Hamiltonian 
off-diagonal matrix elements between 2p2h states and those of original
values.
The dashed curve represents $D_2(\epsilon)$ for the GOE.
}
\label{fig_V22}
\end{figure}

Fig.\ \ref{fig_V22} shows $\lambda_{22}$-dependence of the LSD.
We immediately see that dependence on $\lambda_{22}$ is much weaker 
than that on $\lambda_{12}$.
For $\lambda_{22}<0.2$, the LSD exhibits deviation  from the GOE limit 
for rather wide interval of energy scale, but the deviation itself is
not very large. As $\lambda_{22}$ increases, the LSD approaches the GOE
limit at small energy scales.
(At $\lambda_{22}=0.5$, for instance, the local scaling
dimensions almost follow the GOE when $\epsilon\lesssim10d$.)
For $\lambda_{22}>0.5$ the LSD is not sensitive to the change in $\lambda_{22}$,
and there remains the same deviation from the GOE limit 
even up to the maximal choice of $\lambda_{22}=2.0$.

The insensitivity to $\lambda_{22}$ for $\lambda_{22} > 0.5$
may be understood as follows: 
We first note that 
the energy scale of the mixing among 2p2h states is given
by the spreading width $\gamma_{22}$ of the 2p2h states, and that the mixing
properties are expected to show up in the GOE behaviors for small energy 
scales satisfying $\epsilon < \gamma_{22}$. 
In the case of the original strength
$\lambda_{12}=\lambda_{22}=1$, 
the spreading width $\gamma_{22}^{\rm dm}$ =5.2MeV  of the 2p2h states is
larger than the spreading width $\gamma_{12}^{\rm dm}=1.5$MeV of 
the 1p1h TD states.
If $\lambda_{22}<0.5$, opposite relation $\gamma_{22} <\gamma_{12}$
is realized (n.b. $\gamma_{22}\propto \lambda_{22}^2$ and 
$\gamma_{12}\propto \lambda_{12}^2$), and hence the energy scale where
the deviation from the GOE limit starts is determined by 
$\gamma_{22}$. 
In the case of $\lambda_{22}>0.5$ (including the case of the 
original strength), on the other hand, 
we have $\gamma_{12} <\gamma_{22}$, and the deviation from
the GOE limit is governed by $\gamma_{12}$.
In principle, we can consider a possibility that the LSD exhibits 
the two energy scales $\gamma_{12}$ and $\gamma_{22}$ separately. 
However the deviation related to $\gamma_{22}$ is
small even when $\gamma_{22} < \gamma_{12}$ ($\lambda_{22} < 0.5$), 
making it difficult to extract the energy scale associated 
with $\gamma_{22}$. In the case of the original 
interaction strength, the spreading widths $\gamma_{22}^{\rm dm}=5.2$MeV 
is close to the total width $\gamma_{GR}$ ($\sim$ 5MeV)
of the giant resonance, and thus it is difficult to be probed.
(Notice that we adopted a smoothing width 5MeV.)

Summarizing the above two analyses, the characteristic energy scale of 
the fine structures is related to the spreading width $\gamma_{12}$
of the 1p1h TD states, which is considered as 
the doorway states in the giant resonance decay.

\subsubsection{Isovector strength function}
\label{sec:isovector}

\begin{figure}
\begin{center}
  \begin{minipage}{7.5cm}
       \psfig{file=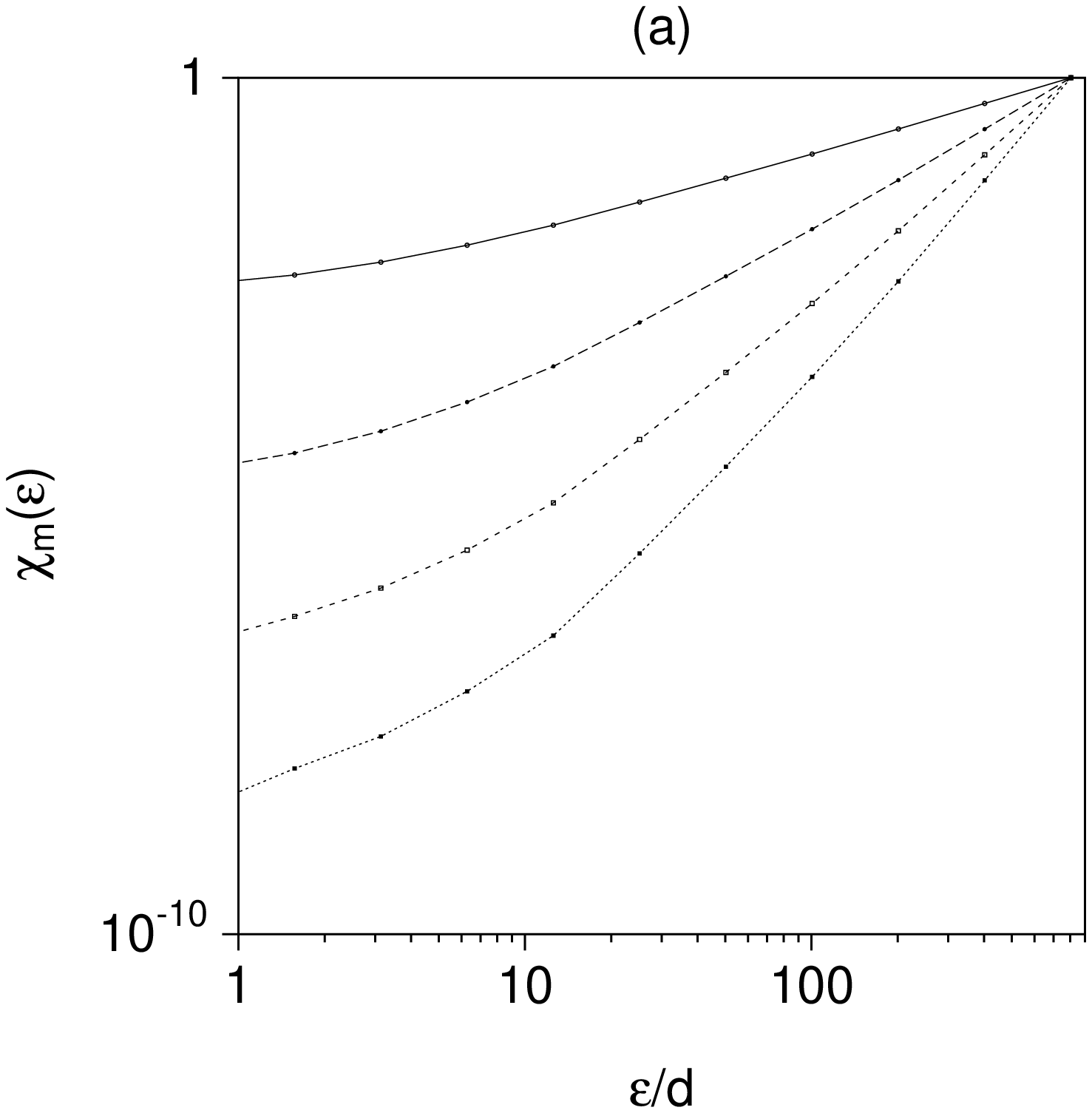,width=7.5cm}
  \end{minipage}
  \begin{minipage}{7.5cm}
       \psfig{file=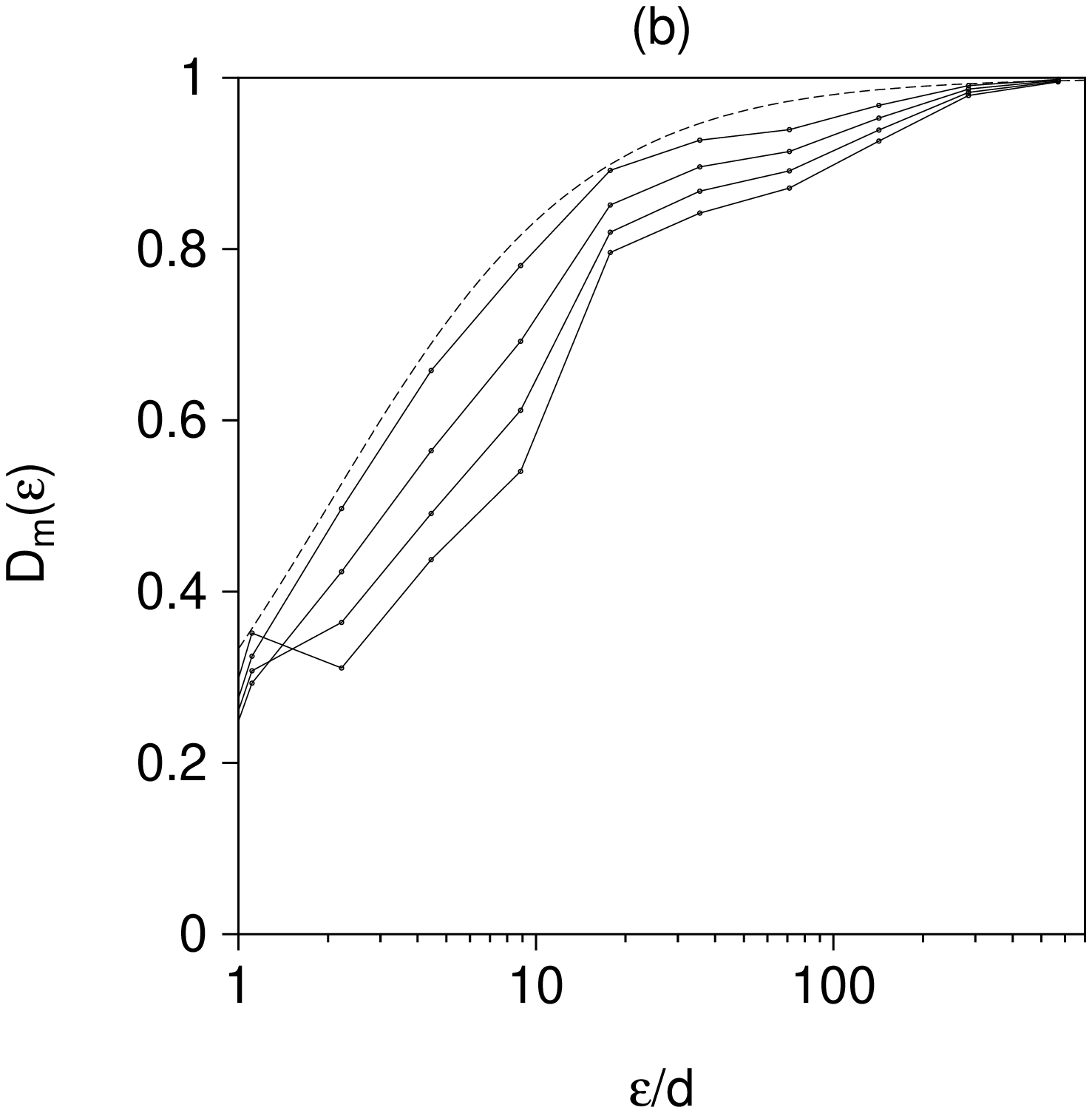,width=7.5cm}
  \end{minipage}
\end{center}
 \caption{
(a)The partition function $\chi_m(\epsilon)$ ($m=2$ to 5)
for the IVGQR corresponding to
the normalized strength funciton plotted in Fig.\ \protect\ref{fig_norstrfun}
(b). (b)Its local scaling dimension $D_m(\epsilon)$ for $m=2$ to 5.
The dashed curve rpresents $D_2(\epsilon)$ for the GOE.
}
\label{fig_lsdiv}
\end{figure}

We have also analyzed the strength distribution for
the isovector quadrupole operator. 
Figures\ \ref{fig_lsdiv}(a) and (b) show the partition
function and the local scaling dimension, respectively, for the isovector
strength function in the energy interval $E=$20-30MeV.
The behavior of the LSD is similar to that of isoscalar strength function.
Namely, the local scaling dimension almost follows the GOE limit
at smaller energy scales whereas the deviation from the GOE limit 
is recognized at intermediate energy scales. 
We also verified that the change of the value of $\epsilon_{\rm f}$ 
is quite sensitive to the change of $\lambda_{12}$ compared to the change
of  $\lambda_{22}$ as in the case of the ISGQR.
Characteristic energy scale of the deviation reads
$\epsilon_{\rm f}\sim 70d \sim 0.9$MeV. This value is smaller than
that for the isoscalar strengths ($\epsilon_{\rm f}\sim 1.7$MeV)
by about a factor of two. Both are consistent with the spreading 
width $\gamma_{12}^{\rm dm}=1.5$MeV. 


\section{Conclusions}
\label{sec:conclusion}

We have analyzed fine structures of the giant 
quadrupole resonances described by means of the microscopic
shell model including up to 2p2h configurations. By applying the
scaling analysis of strength fluctuations that utilizes
the local scaling dimension, we extracted the
energy scale $\epsilon_{\rm f}=1.7$ MeV for the ISGQR and
$\epsilon_{\rm f}=0.9$ MeV for the IVGQR that characterize 
the fine structure.
We discussed the
origin of the characteristic energy scale in terms of the
spreading widths of the 1p1h and 2p2h states. We found a clear
correlation between the characteristic energy scale and 
the spreading width $\gamma_{12}$ of the 1p1h Tamm-Dancoff states,
which play a role of doorway states in the present shell model
description of the giant quadrupole resonances in $^{40}$Ca.

In the present paper, we intended to illustrate how the
scaling analysis can be utilized to investigate 
the fine structures and their origin.
Although the adopted model has some realistic features,
we do not claim that the present shell model
predicts all the quantitative features of the quadrupole
strength distributions since the particle escaping width for instance is
neglected in the present model.
Also, other mechanisms such as the surface phonon
coupling are not explicitly taken into account in the
present model. Apparently, one may need more realistic model 
if one intends to make 
a quantitative prediction or comparison with the experiment.
Instead we would like to emphasize that 
the local scaling analysis provides a general tool for quantitative
investigation of the fine structures, and 
the scaling analysis itself 
can be applied to any kinds of models which exhibit fine structures
in the strength functions.

\end{document}